\documentclass[aps,prl,twocolumn,superscriptaddress,showpacs,floatfix]{revtex4-2}

\usepackage{amsmath,amssymb,amsfonts,float,graphics,epsfig,epstopdf,color,verbatim,tabularx,bm,multirow,appendix}
\usepackage[utf8]{inputenc}
\usepackage[T1]{fontenc}
\usepackage{xcolor}
\usepackage{cancel}

\usepackage{dsfont}
\usepackage{textcomp}
\usepackage{yfonts}
\usepackage{bm}
\usepackage{subfigure}
\usepackage{mathrsfs}
\usepackage{graphicx}
\usepackage{verbatim}
\usepackage{hyperref}
\usepackage{multirow}
\usepackage{braket}
\usepackage[normalem]{ulem}
\usepackage{tikz}
\usetikzlibrary{calc}
\usetikzlibrary{shapes.multipart}

\newcommand{\pdiff}[2]{\frac{\partial #1}{\partial #2}}

\newcommand{\ua}{\uparrow}
\newcommand{\da}{\downarrow}

\newcommand{\ra}{\rightarrow}
\renewcommand{\vec}[1]{{\mathbf #1}}

\newcommand{\comments}[1]{}

\newcommand{\stkout}[1]{\ifmmode\text{\sout{\ensuremath{#1}}}\else\sout{#1}\fi}

\makeatletter
\def\l@subsubsection#1#2{}
\makeatother

\begin{document}

\title{Emus live on the chiral Gross-Neveu-Yukawa archipelago}

\author{Ting-Tung Wang}
\affiliation{Department of Physics and HKU-UCAS Joint Institute of Theoretical and Computational Physics, The University of Hong Kong, Pokfulam Road, Hong Kong SAR, China}

\author{Zi Yang Meng}
\email{zymeng@hku.hk}
\affiliation{Department of Physics and HKU-UCAS Joint Institute of Theoretical and Computational Physics, The University of Hong Kong, Pokfulam Road, Hong Kong SAR, China}

\begin{abstract}
It is expected that the Gross-Neveu-Yukawa (GNY) chiral Ising transition of Dirac fermions coupled with scalar field in (2+1)D will be the first fermionic quantum critical point that various methods such as conformal bootstrap~\cite{Erramilli-Gross-2023}, perturbative renormalization group~\cite{Zerf-Four-loop-2017} and quantum Monte Carlo (QMC) simulations~\cite{Liu-Designer-2020}, would yield the converged critical exponents -- serving the same role as the Ising and $O(N)$ models in the textbooks of statistical and quantum physics. However, such expectation has not been fully realized from the lattice QMC simulations due to the obstacles introduced by the UV finite size effect. In this work, by means of the elective-momentum ultra-size (EMUS) QMC method~\cite{Liu-Elective-2019}, we compute the critical exponents of the $O(N/2)^2 \rtimes \mathbb{Z}_2$ GNY $N=8$ chiral Ising transition on a 2D $\pi$-flux fermion lattice model between Dirac semimetal and quantum spin Hall insulator phases. With the matching of fermionic and bosonic momentum transfer and collective update in momentum space, our QMC results provide the fully consistent exponents with those obtained from the bootstrap and perturbative approaches. In this way, the Emus now live happily on the $N=8$ island and could explore the chiral Gross-Neveu-Yukawa archipelago~\cite{Erramilli-Gross-2023} with ease. 
\end{abstract}

\date{\today}
\maketitle

\noindent{\textcolor{blue}{\it Introduction.}---} 
Just like the Ising and $O(N)$ models are the simplest $(2+1)$D universality classes that the perturbative renormalization group (RG) analysis ($\epsilon$-expansion)~\cite{litimIsing2011,kompanietsMinimally2017}, the conformal bootstrap~\cite{showkSolving2012,kosPrecision2016,chesterCarving2020,chesterBootstrapping2021} and lattice model simulations~\cite{hasenbuschFinite2010} have provided highly consistent and well-converged results -- serving as the textbook example for development of many-body methodologies, the simplest $(2+1)$D universality class involving fermions -- the Gross-Neveu-Yukawa (GNY) model of Dirac fermions coupled with  scalar bosonic field -- is expected to provide the similar level of consistency and bring our understanding of the quantum phase transitions in interacting Dirac fermion systems to more solid ground. 

Such consistency not only has theoretical impact towards quantum field theory and high-energy physics~\cite{graceyThreeloop1990,Erramilli-Gross-2023,graceyFour2016,Zerf-Four-loop-2017}, but is also intimately related to the on-going research in D-wave superconductor and nematic quantum criticality~\cite{vojtaQuantum2000,schwabNematic2022}, graphene~\cite{geimRise2007,herbutInteractions2006,herbutTheory2009,herbutRelativistic2009,Zerf-Four-loop-2017}, twisted bilayer graphene~\cite{caoUnconventional2018,caoCorrelated2018} and other quantum moir\'e materials~\cite{kennesMoire2021,andreiMarvels2021}, as well as the kagome metallic systems~\cite{yinTopological2022,kangDirac2020}, where the transition from Dirac and Weyl semimetals (with the fermion flavor tuned by the spin, valley and layer degrees of freedom and the Coulomb interaction tuned by gating and twist angles) to various symmetry-breaking phases holds the key to understand the intriguing phenomena therein.
\begin{table}[htp!]
	\begin{tabular}{ |c | c c c| } 
		\hline
		& $1/\nu$ & $\eta_\phi$ & $\eta_\psi$ \\
		\hline
		This work & 1.07(12) & 0.72(6) & 0.04(2) \\ 
		Previous QMC \cite{Liu-Designer-2020}& 1.0(1) & 0.59(2) & 0.05(2) \\ 
		Previous QMC \cite{Chandrasekharan-Quanum-2013} & 1.20(1) & 0.62(1) & 0.38(1) \\
		$\epsilon$-expansion \cite{Ihrig-Critical-2018}& 0.993(27) & 0.704(15)& 0.043(12)   \\ 
		Conformal bootstrap \cite{Erramilli-Gross-2023}& 0.998(12) & 0.7329(27) & 0.04238(11) \\ 
		\hline
	\end{tabular}
	\caption{\textbf{Emus live on the chiral Gross-Neveu-Yukawa archipelago.} This table summarizes the critical exponents of $N=8$ chiral Ising GNY transition. We compare the EMUS-QMC results with previous QMC results and those from the latest $\epsilon$-expansion and bootstrap estimates. The bootstrap estimates are obtained for the three external operators $\Delta_{\epsilon}=3-\frac{1}{\nu}$, $\Delta_{\sigma}=\frac{1+\eta_{\phi}}{2}$ and $\Delta_{\psi}=\frac{1+\eta_\psi}{2}$ of GNY island with $O(8)$ global symmetry~\cite{Erramilli-Gross-2023}. The $\epsilon$-expansion work~\cite{Ihrig-Critical-2018} relies on the DREG3 prescription to analytically continue spinors away from $d = 4$.}
	\label{table:1}
\end{table}

However, such consistency, especially in the form of the critical exponents or the scaling dimensions of external operators in the CFT data, has not been fully reached for the simplest one -- the GNY chiral Ising transition with $O(N/2)^2 \rtimes \mathbb{Z}_2$ global symmetry -- for now. The present $\epsilon$-expansion~\cite{graceyFour2016,Zerf-Four-loop-2017,Ihrig-Critical-2018}, conformal bootstrap with $O(N)$ global symmetry~\cite{Erramilli-Gross-2023,Iliesiu-Bootstrapping-2018}  and lattice model QMC simulation~\cite{Chandrasekharan-Quanum-2013,Liu-Elective-2019,He-Dynamical-2018,Liu-Designer-2020,Tabatabaei-Chiral-2022,Bonati-Chiral-2023} are giving rise to closer exponents over the years (see Tab.~\ref{table:1}), except for the remaining boson anomalous dimension exponent $\eta_\phi$. The $\sim 20\%$ deviation from the latest QMC study~\cite{Chandrasekharan-Quanum-2013,Liu-Designer-2020} compared with that from $\epsilon$-expansion~\cite{Zerf-Four-loop-2017,Ihrig-Critical-2018} and conformal bootstrap~\cite{Erramilli-Gross-2023}, comes from the fact that, although in the latest lattice model simulation~\cite{Liu-Designer-2020}, the critical bosonic and fermionic modes are designed with the same velocity at the bare level, the actually coupled system when driving to the quantum critical point, still acquires different velocities of the critical modes at the finite size studied (see Fig.1 in Ref.~\cite{Liu-Designer-2020}), and it has been observed that such difference at the UV is sufficient to cause significant drifts of the exponents in the finite size analyses (see Figs.2 and 3 in Ref.~\cite{Liu-Designer-2020}) and renders the access of the thermodynamic limit difficult. From these experience, one sees that better model design and algorithmic developments, {\it with less computation time and human time}, are critically needed to overcome the problem and bring the consistent results with bootstrap and $\epsilon$-expansion for GNY chiral Ising universality classes.

\begin{figure*}[htp!]
	\includegraphics[width=0.9\textwidth]{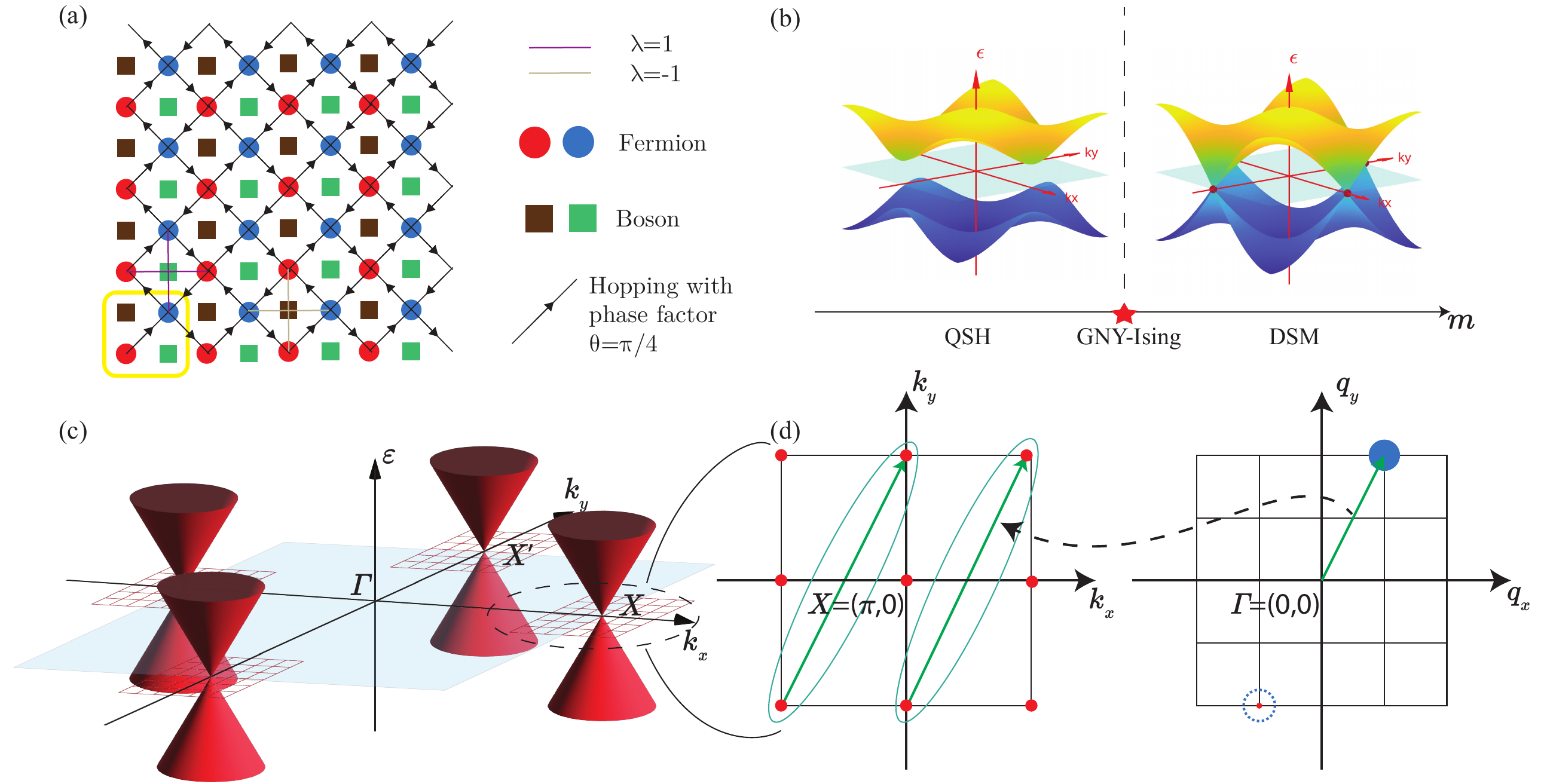}
	\centering
	\caption{\textbf{Chiral-Ising GNY Model and EMUS algorithm.} (a) The lattice model of $N=8$ chiral Ising GNY in real space. Each unit cell (yellow square) contains 2 boson and 2 fermion sites. Black arrows indicate fermionic hopping with a phase factor $\theta=\frac{\pi}{4}$, which generates a $\pi$-flux in each plaquette and lead to two Dirac cones in the Brillouin zone (BZ). Purple and beige lines indicate coupling terms with strength equals to the bosonic field multiplied with $\lambda=\pm1$, depending on the boson sub-lattice. (b) Fermion dispersions. When $m$ is small (large), bosonic fields are in ferromagnetic (paramagnetic) phases and the fermions are in the massive QSH (massless DSM) phases, separated by the GNY chiral Ising transition at $m_c$. (c) In momentum space, two patches near Dirac points (denoted by the red grids) are simulated in EMUS-QMC. (d) An example of EMUS update scheme with patch size $L_f=2$ (corresponds to $L=12$ in the original model). The left panel is one of the patches in fermion BZ. The right panel is the allowed momentum transfer for bosons.}
	\label{fig:1}
\end{figure*}
In this work, we achieve this goal by means of the elective-momentum ultra-size (EMUS)-QMC method~\cite{Liu-Elective-2019,xuRevealing2019,liuItinerant2019}, we compute the critical exponents of the $N=8$ chiral Ising  GNY transition on a 2D $\pi$-flux fermion lattice model between Dirac semimetal (DSM) and quantum spin Hall insulator (QSH) phases~\cite{He-Dynamical-2018,Liu-Designer-2020}. By designing the matching of fermionic and bosonic momentum transfer within the high resolution patches in the Brillouin zone (BZ) and collective update completely in the momentum space, we have effectively accessed much larger system sizes and better data quality {\it with reduced computational cost as well as human time}. Our QMC results yield the crossing of the RG-invariant ratio with very small drifts and our stochastic finite size analysis finds fully controlled exponents (see Tab.~\ref{table:1}), finally in agreement with those obtained from the bootstrap and $\epsilon$-expansion. 

With our new computation protocol, the Emus now live happily on the $N=8$ island of the chiral GNY archipelago~\cite{Erramilli-Gross-2023}, and they can readily jump to other islands with simple change of the simulation code and further explore the exciting and vast ocean of CFTs. Relevance towards the experiment on interacting Dirac fermion systems is also discussed.

\begin{figure*}[htp!]
\centering
\includegraphics[width=\textwidth]{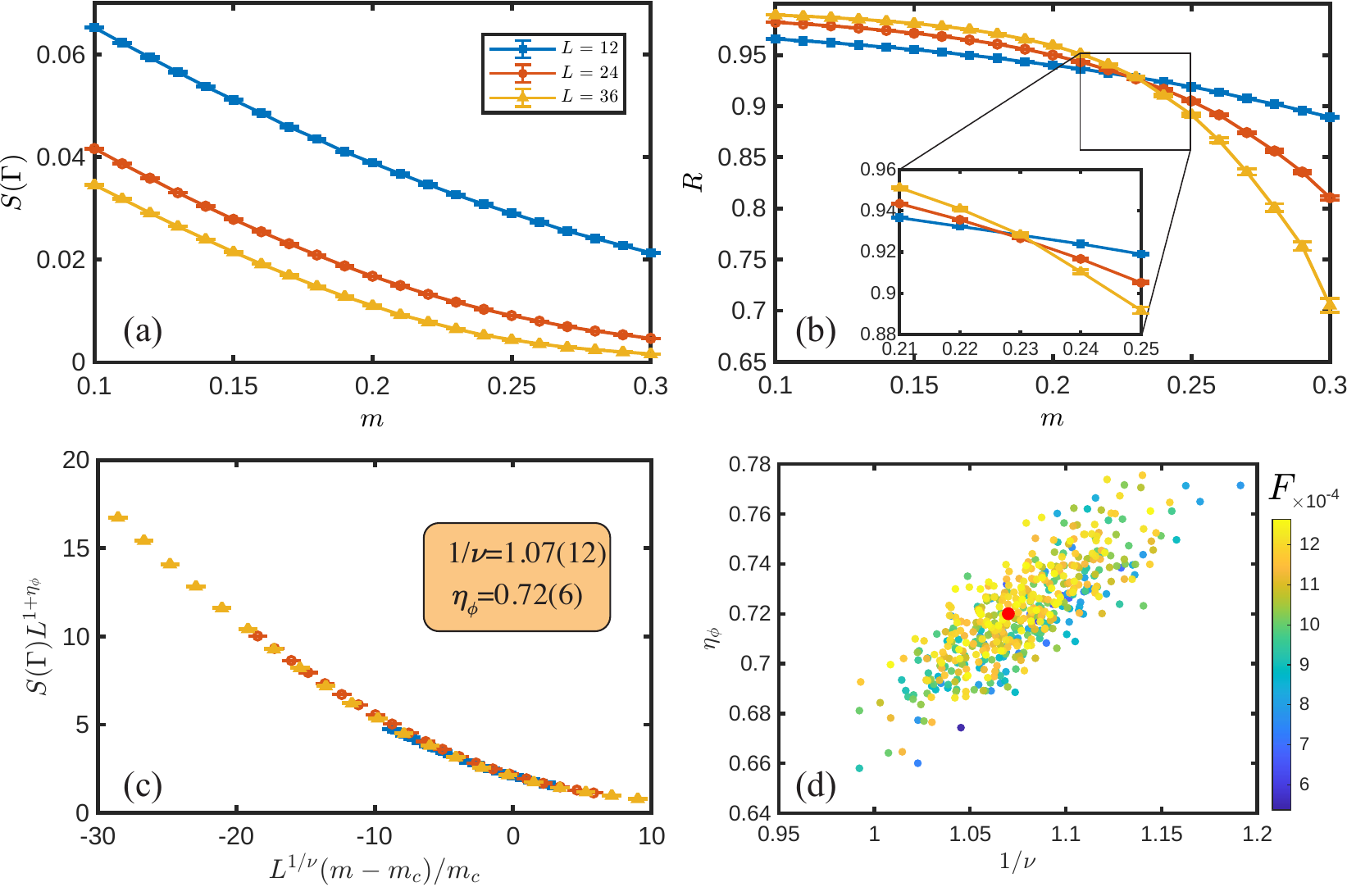}
\caption{\textbf{$N=8$ chiral Ising GNY CFT data.} (a) The square of the bosonic order parameter $S(\Gamma)$, and (b) the correlation ratio $R$ against the control parameter $m$ for system size $L=12, 24, 36$. (c) and (d) The data collapse of $S(\Gamma)$ versus $m$ and the stochastic analysis of the error bound of $\{1/\nu, \eta_\phi\}$. Results yield $1/\nu=1.07(12)$ and the bosonic anomalous dimension $\eta_\phi=0.72(6)$. The colorbar in (d) denotes the $F$ ratio of the goodness of fit. Red dot is the set of exponents used in the collapse in (c).}
	\label{fig:2}
\end{figure*}

\noindent{\textcolor{blue}{\it The chiral GNY Model.}---} 
At the level of field theory, our chiral GNY model describes the situation of $N/4$ four-component Dirac interacting with a bosonic scalar field $\phi$. The coefficient $m$ in the $m\phi^2$ term of bosonic Lagrangian (see Eq.~\eqref{eq:eq1}) has a critical value $m_c$ below which the bosonic scalar field spontaneously acquires a finite expectation value, giving a mass to the fermions (the QSH phase in Fig.~\ref{fig:1} (b)). Above $m_c$, the expectation value of $\phi$ vanishes, and the fermions go back to the massless form (the DSM phase in Fig.~\ref{fig:1} (b)). At $m_c$, the system is expected to flow to the chiral GNY CFT. %As discussed in the introduction, the GNY models have been intensively studied previously with the conformal bootstrap~\cite{iliesiuBootstrapping2016,Iliesiu-Bootstrapping-2018,Erramilli-Gross-2023}, the resulting bounds exhibited a sequence of islands (the GNY archipelago) on the boundary of the space of allowed CFT data, which showed good agreement with perturbative estimates of the scaling dimensions~\cite{Ihrig-Critical-2018,graceyFour2016,Zerf-Four-loop-2017} and large-$N$ expansions~\cite{iliesiuBootstrapping2016,graceyAnomalous1992,derkachovCalculation1993,petkouOperator1996}, as well as large-scale lattice QMC simulations~\cite{Chandrasekharan-Quanum-2013,Liu-Elective-2019,He-Dynamical-2018,Liu-Designer-2020,Tabatabaei-Chiral-2022,Bonati-Chiral-2023}. 
We note in the literatures there are two different GNY models with the same number of fermions but different global symmetry groups. In addition to the chiral GNY model with $O(N/2)^2 \rtimes \mathbb{Z}_2$ global symmetry in this work, there is also the GNY model with $O(N)$ global symmetry investigated in the recent bootstrap work~\cite{Erramilli-Gross-2023}. However, since these models are nearly degenerate and being only distinguishable at high
perturbative order, the differences between the scaling dimensions of external operators $\{ \Delta_\psi=\frac{1+\eta_\psi}{2}, \Delta_\sigma=\frac{1+\eta_{\phi}}{2}, \Delta_{\epsilon}=3-\frac{1}{\nu}\}$ are very small ($\sim 3\times10^{-6}$) for $N=1,2,4,8$~\cite{Erramilli-Gross-2023,Zerf-Four-loop-2017}, we hence focus on the lattice realization of $N=8$ chiral Ising GNY with $O(N/2)^2 \rtimes \mathbb{Z}_2$ global symmetry.
%However, as shown in the summary in Tab.~\ref{table:1}, although the obtained exponents are converging over the years, for the correlation length exponent $\nu$ and the fermion anomalous dimension $\eta_\psi$, there still exists an obvious discrepancy of the bosonic anomalous dimension $\eta_\phi \sim 0.6$ from previous QMC and $\eta_\phi \sim 0.7$ from the latest $\epsilon$-expansion and bootstrap, beyond the numerical uncertainties. In this work, we would like to explain the reason behind this last discrepancy and demonstrate how to overcome it with better model design and algorithmic development from the EMUS-QMC method, which actually takes less computation and human time.

We adapt the lattice model shown in Fig.~\ref{fig:1} (a) with the Lagrangian~\cite{Liu-Designer-2020}
\begin{equation}
\mathscr{L}=\mathscr{L}_{Fermion}+\mathscr{L}_{Boson}+\mathscr{L}_{Coupling}
\label{eq:eq1}
\end{equation}
where
\begin{align*}
&\mathscr{L}_{Fermion}=\sum_{\langle i,j\rangle\sigma}\psi^\dagger_{i,\sigma}[(i\partial_\tau-\mu)\delta_{ij}-te^{i\sigma\theta_{ij}}]\psi_{j,\sigma}+h.c.,\\
&\mathscr{L}_{Boson}=\sum_i[\frac{1}{4}(\pdiff{\phi_i}{\tau})^2+m\phi_i^2+\phi_i^4]+\sum_{(i,j)}J_{ij}(\phi_i-\phi_j)^2,\\
&\mathscr{L}_{Coupling}=\sum_{\langle\langle i,j\rangle\rangle,\sigma}\lambda_{i,j}\phi_{p}\psi^\dagger_{i,\sigma}\psi_{j,\sigma}+h.c.,
\end{align*}
and $\psi_{i,\sigma}$ is the fermionic operator on each site $i$ with spin ${\sigma=\ua,\da}$. $\mathscr{L}_{Fermion}$ describes fermions with nearest neighbor hopping $t=1$ and a phase factor $\theta=\pi/4$ for each bond, which introduces a $\pi$-flux in each plaquette and leads to two Dirac cones at $X=(0,\pi)$ and $(\pi,0)$ in the BZ. We set $\mu=0$ to ensure the half-filling of fermions.

In $\mathscr{L}_{Boson}$, $\phi_i$ is the scalar bosonic field, $m$ is the mass term to tune the boson across the chiral GNY Ising transition. The imaginary time derivative term provides the quantum fluctuations. $J_{ij}$ are interaction terms up to forth-nearest neighbor, and their magnitudes are set to ${J_1=4t^2/5}$, ${J_2=-J_1/8}$, ${J_3=J_1/63}$, and ${J_4=-J_1/896}$. This combination ensures the largest linear region in bosonic dispersion, and the bosonic velocity $v_b$ is equal to that of the bare fermion $v_f=v_b=\sqrt{2}t$~\cite{Liu-Designer-2020}.
%When tuning $m$ up, the ${m\phi^2}$ term will suppress the ferromagnetic order of the scalar field, and hence drives the bosonic field through a quantum phase transition of $(2+1)$D Ising type, going from ferromagnetic phase to paramagnetic phase.

$\mathscr{L}_{Coupling}$ is a next nearest neighbor hopping for fermion, whose strength is determined by the boson $\phi_p$ sitting on the bond and the difference of two boson sub-lattices $\lambda_{i,j}=\pm 1$. In the symmetry-breaking phase of bosonic field, this will open a gap at the Dirac points in fermion dispersion and transform the DSM to QSH insulator~\cite{He-Dynamical-2018}.

\noindent{\textcolor{blue}{\it EMUS-QMC method.}---} 
As shown in Fig.~\ref{fig:1} (b), the linear dispersion of Dirac cones only occupies a small part of the BZ and simulating the region out of linear dispersion will not contribute much to the critical behaviour. In previous studies, different ways are introduced to bypass this issue. One can write the model with a single Dirac cone (the SLAC fermion) covering most of the BZ~\cite{langQuantum2019,Tabatabaei-Chiral-2022}. However, although SLAC fermion avoids the Nielsen-Ninomiya theorem~\cite{nielsenAbsenceI1981,nielsenAbsenceII1981}, one has to pay the price of long-range hoppings and the violation of locality has been shown to fundamentally change the associated universality class~\cite{Koziol-Quantum-2021,liaoCaution2022,wangValidity2022}. One can also enlarge the linear dispersion region without violation of the locality, by adding longer neighbor hoppings with appropriate strength and set the bare velocities $v_b=v_f$~\cite{Liu-Designer-2020}. However, as discussed in the introduction, one still experience strong finite size effect when tuning the system to the interacting fixed point, and the drift in the crossings of RG-invariant ratios is very strong such that it gives rise to the $\sim 20\%$ deviation of the bosonic anomalous dimension $\eta_{\phi}$ in the latest QMC and the $\epsilon$-expansion and conformal bootstrap, as shown in Tab.~\ref{table:1}.

Facing with these difficulties, here we work in a different direction, instead of enlarging the linear region, we ONLY simulate the linear region -- the momentum space near Dirac points -- with the EMUS-QMC method~\cite{Liu-Elective-2019,xuRevealing2019,liuItinerant2019}. The EMUS scheme is different from the usual fermion determinant QMC, in that, instead of simulating the lattice model with a homogeneous $L\times L$ grid in real (and momentum) space, it focus on small patches of BZ that are important in the IR limit and ignore the momenta far from them. In the present model, the "hot spots" are the two Dirac points at $X$($X'$). Although the hard cutoff in momentum space of EMUS effectively renders the QMC to simulate a different model -- whose dispersion near the Dirac points is same as the original one -- with different non-universal observables, such as magnitude of order parameter, location of phase transition point etc., we find the CFT data (critical properties) of the EMUS simulation share the same IR structure with the original ones, as shown in this work and our previous examples~\cite{Liu-Elective-2019,xuRevealing2019,liuItinerant2019}. We leave the detailed description of the EMUS-QMC to the Supplementary Material (SM)~\cite{suppl}, but just to highlight that we can now simulate system sizes up to $36\times36$, with {\it with less computation time and human time}, compared with the $16\times16$ in traditional method~\cite{Liu-Designer-2020}.

\begin{figure}[htp!]
\includegraphics[width=\columnwidth]{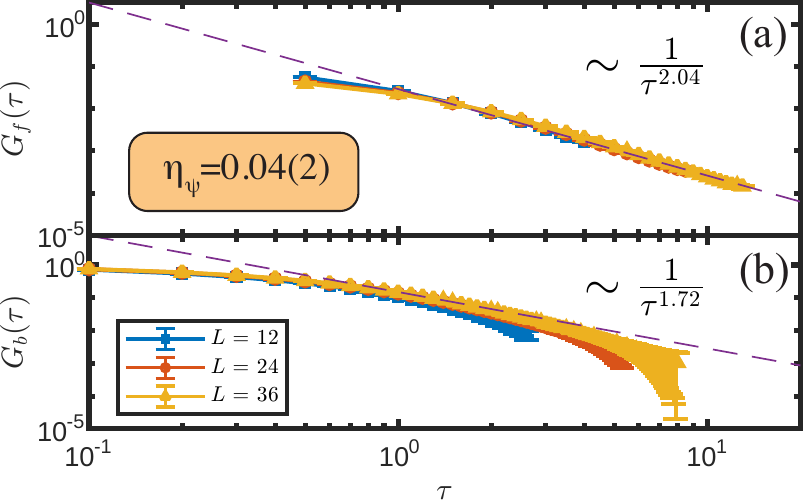}
\centering
\caption{\textbf{Fermion and Boson Green's functions at the chiral GNY CFT.} We fit the imaginary time decays of fermionic (a) and bosonic (b) Green's functions, to extract the anomalous dimension $\eta_\psi=0.04(2)$ and verify the consistency of $\eta_\phi=0.72$ obtained from Fig.~\ref{fig:2}.}
	\label{fig:3}
\end{figure}

\noindent{\textcolor{blue}{\it Results.}---} We perform EMUS-QMC on Eq.~\eqref{eq:eq1} with patch size $L_f=2, 4, 6$, which is equivalent to the original system size $L=12, 24, 36$ and set the inverse temperature $\beta=\frac{3}{4}L$. We compute the magnetic structure factor of the bosonic field $S(\vec{q})=\frac{1}{L^2}\langle\phi_{\vec{q}}\phi_{-\vec{q}}\rangle=\frac{1}{L^4}\sum_{i,j}e^{-i\vec{q}\cdot(\vec{r}_i-\vec{r}_j)}\langle\phi_i\phi_j\rangle$, and use the $S(\Gamma)$ (the square of the order parameter) and the RG-invariant correlation ratio $R=1-\frac{S(\Gamma+\Delta\vec{q})}{S(\Gamma)}$~\cite{pujariInteraction2016}, where $\Delta{\mathbf{q}}=\frac{2\pi}{L}(0,1) \text{ or } (1,0)$, to extract the GNY critical exponents. Since we work directly in momentum space, the Fourier components $\phi_{\mathbf{q}}$ can be used in measurement on the fly, which is faster than the same measurement in real-space QMC.

One sees in Fig.~\ref{fig:2} (b) that the drift of crossing points of $R$ is extremely small compare to that in Fig.2 of Ref. \cite{Liu-Designer-2020}. This is a strong indication that our simulation has smaller finite size effect, as we attain larger effective system sizes. With the obtained $S(\Gamma)$ and $R$ in Fig.~\ref{fig:2} (a) and (b), we further perform a stochastic data collapse in Fig.~\ref{fig:2} (c) and (d), to unbiasedly determine the optimized critical exponents $\{1/\nu,\eta_\phi\}$. The detailed description of such analysis is given in SM~\cite{suppl}, and here we outline the main procedure and results. 

We sample the values of $\{1/\nu,\eta_\phi\}$ from 2D parameter space in Fig.~\ref{fig:2} (d). Whenever a set of critical exponents is proposed, we try to fit a single curve with rescaled data from all system sizes (as in Fig.~\ref{fig:2} (c)), and a ratio $F$ is computed to determined the goodness of fit. Then we continue varying the exponents to find where $F$ attains its minimum. This whole process is repeated 1000 times with random noise added to $S(\Gamma)$ within its error bar and with random initial guesses in the $\{1/\nu,\eta_\phi\}$ space, so as to estimate the error bound for the optimal exponents, which is shown in Fig.~\ref{fig:2} (d). The color of each dot indicates its magnitude of convergent $F$ ratio, where those of the blue ones are lower, and hence fit better. In this way, the $1/\nu=1.07(12)$ and $\eta_\phi=0.72(6)$ are obtained. The red dot indicates the exponents generated from the mean value of $S(\Gamma)$ without noise, which collapses well as shown in Fig.~\ref{fig:2} (c). 

To obtain the fermion anomalous dimension $\eta_\psi$. We monitor the imaginary time Green's function by utilizing the Lorentz symmetry at the GNY-Ising CFT. Fig.~\ref{fig:3} shows the imaginary time decay of the fermionic and bosonic Green's function at the critical point $m_c$, where $G_f(\tau)=\frac{1}{L^2}\sum_{\vec{k}}\langle \psi_{\vec{k}}(\tau)\psi^\dagger_{\vec{k}}(0)\rangle$, and $G_b(\tau)=\frac{1}{L^2}\sum_{\vec{q}}\langle \phi_{\vec{q}}(\tau)\phi_{-\vec{q}}(0)\rangle$, respectively. Both panels are in log-log scale, the dotted straight lines indicate power law decay of both correlation functions. From Fig.~\ref{fig:3} (a), we can extract the anomalous dimension of fermion $\eta_\psi=0.04(2)$, by the relation $G_f(\tau)\propto1/\tau^{1+\eta_\psi}$. While in Fig.~\ref{fig:3} (b), a straight line of $G_b(\tau) \propto 1/\tau^{1.72}$ is drawn, using the $\eta_\phi=0.72$ just obtained. One can see both straight lines match well with the data as system size $L$ increases, which reflects the robustness of $\eta_\psi$ and $\eta_\phi$.

\noindent{\textcolor{blue}{\it Discussion.}---} Tab.~\ref{table:1} summarizes our results and the previous ones from QMC, $\epsilon$-expansion and conformal bootstrap. One sees the textbook level consistency of the $\{1/\nu, \eta_\phi, \eta_\psi\}$ for the $N=8$ chiral Ising GNY with the conformal bootstrap~\cite{Erramilli-Gross-2023} and $\epsilon$-expansion~\cite{Ihrig-Critical-2018}, is finally achieved. We note that the present EMUS-QMC simulation actually consume much less computation resource compared with our previous one~\cite{Liu-Designer-2020},
%However, we are not able to extract the correction exponent $\omega$, which is formerly done in QMC, by extrapolating crossing points of correlation ratio $R$ of different system size. In our case, we only have 3 available system size and 2 crossing points. Therefore, such extrapolating will be extremely ill-conditioned. One can try to attain some intermediate system sizes by  changing the ratio between $L$ and $L_f$. However, this will leads to yet another effective model, and thus need to be consider separately, although they share the same set of critical exponents. In other words, some new degrees of freedom are introduced, and new coefficients need to be found.
and it can be readily extended to other chiral GNY islands with different number of fermion flavours and symmetries of the bosonic field, by simply changing the power in the fermion determinant and the form of boson energy difference~\cite{suppl}.

The EMUS-QMC can also be used to investigate other GNY transitions from DSM to plaquette valence bond solids~\cite{satoDirac2017,ouyangProjection2021,liaoDiracI2022,liaoDiracII2022,liaoDiracIII2022}, $SU(2)$ QSH~\cite{liuSuperconductivity2019,wangDoping2021}, nematic order~\cite{schwabNematic2022}, inter-valley coherent and valley polarized orders~\cite{liaoValence2019,liaoCorrelation2021,liaoCorrelationReview2021,panDynamical2022,panSport2022} as well as superconductivity~\cite{xuCompeting2021,zhangSuperconductivity2022}, in graphene and twisted bilayer graphene and kagome metal systems. We foresee Emus will soon explore the exciting islands in the vast ocean of CFTs and the broad continent of the realistic 2D quantum materials.

\noindent{\it Acknowledgments}\,---\, 
We thank Junchen Rong, Ning Su for inspiring discussions on the GNY transition from conformal bootstrap perspective, Michael Scherer on the enlightening education on the GNY transition from perturbative RG aspects and Gaopei Pan and Weilun Jiang for useful technical suggestions on EMUS-QMC implementation. TTW and ZYM acknowledge support from the RGC of Hong Kong SAR of China (Project Nos. 17301420, 17301721, AoE/P-701/20, 17309822,  HKU C7037-22G), the ANR/RGC Joint Research Scheme sponsored by Research Grants Council of Hong Kong SAR of China and French National Research Agency(Project No. A\_HKU703/22) and the Seed Funding “Quantum-Inspired explainable-AI” at the HKU-TCL Joint Research Centre for Artificial Intelligence.
The authors also acknowledge the HPC2021 system under the Information Technology Services and the Blackbody HPC system at the Department of Physics, University of Hong Kong for providing HPC resources.

\bibliography{GNY-EQMC}

\begin{thebibliography}{60}%
\makeatletter
\providecommand \@ifxundefined [1]{%
 \@ifx{#1\undefined}
}%
\providecommand \@ifnum [1]{%
 \ifnum #1\expandafter \@firstoftwo
 \else \expandafter \@secondoftwo
 \fi
}%
\providecommand \@ifx [1]{%
 \ifx #1\expandafter \@firstoftwo
 \else \expandafter \@secondoftwo
 \fi
}%
\providecommand \natexlab [1]{#1}%
\providecommand \enquote  [1]{``#1''}%
\providecommand \bibnamefont  [1]{#1}%
\providecommand \bibfnamefont [1]{#1}%
\providecommand \citenamefont [1]{#1}%
\providecommand \href@noop [0]{\@secondoftwo}%
\providecommand \href [0]{\begingroup \@sanitize@url \@href}%
\providecommand \@href[1]{\@@startlink{#1}\@@href}%
\providecommand \@@href[1]{\endgroup#1\@@endlink}%
\providecommand \@sanitize@url [0]{\catcode `\\12\catcode `\$12\catcode
  `\&12\catcode `\#12\catcode `\^12\catcode `\_12\catcode `\%12\relax}%
\providecommand \@@startlink[1]{}%
\providecommand \@@endlink[0]{}%
\providecommand \url  [0]{\begingroup\@sanitize@url \@url }%
\providecommand \@url [1]{\endgroup\@href {#1}{\urlprefix }}%
\providecommand \urlprefix  [0]{URL }%
\providecommand \Eprint [0]{\href }%
\providecommand \doibase [0]{https://doi.org/}%
\providecommand \selectlanguage [0]{\@gobble}%
\providecommand \bibinfo  [0]{\@secondoftwo}%
\providecommand \bibfield  [0]{\@secondoftwo}%
\providecommand \translation [1]{[#1]}%
\providecommand \BibitemOpen [0]{}%
\providecommand \bibitemStop [0]{}%
\providecommand \bibitemNoStop [0]{.\EOS\space}%
\providecommand \EOS [0]{\spacefactor3000\relax}%
\providecommand \BibitemShut  [1]{\csname bibitem#1\endcsname}%
\let\auto@bib@innerbib\@empty
%</preamble>
\bibitem [{\citenamefont {Erramilli}\ \emph {et~al.}(2023)\citenamefont
  {Erramilli}, \citenamefont {Iliesiu}, \citenamefont {Kravchuk}, \citenamefont
  {Liu}, \citenamefont {Poland},\ and\ \citenamefont
  {Simmons-Duffin}}]{Erramilli-Gross-2023}%
  \BibitemOpen
  \bibfield  {author} {\bibinfo {author} {\bibfnamefont {R.~S.}\ \bibnamefont
  {Erramilli}}, \bibinfo {author} {\bibfnamefont {L.~V.}\ \bibnamefont
  {Iliesiu}}, \bibinfo {author} {\bibfnamefont {P.}~\bibnamefont {Kravchuk}},
  \bibinfo {author} {\bibfnamefont {A.}~\bibnamefont {Liu}}, \bibinfo {author}
  {\bibfnamefont {D.}~\bibnamefont {Poland}},\ and\ \bibinfo {author}
  {\bibfnamefont {D.}~\bibnamefont {Simmons-Duffin}},\ }\href
  {https://doi.org/10.1007/JHEP02(2023)036} {\bibfield  {journal} {\bibinfo
  {journal} {Journal of High Energy Physics}\ }\textbf {\bibinfo {volume}
  {2023}},\ \bibinfo {pages} {36} (\bibinfo {year} {2023})}\BibitemShut
  {NoStop}%
\bibitem [{\citenamefont {Zerf}\ \emph {et~al.}(2017)\citenamefont {Zerf},
  \citenamefont {Mihaila}, \citenamefont {Marquard}, \citenamefont {Herbut},\
  and\ \citenamefont {Scherer}}]{Zerf-Four-loop-2017}%
  \BibitemOpen
  \bibfield  {author} {\bibinfo {author} {\bibfnamefont {N.}~\bibnamefont
  {Zerf}}, \bibinfo {author} {\bibfnamefont {L.~N.}\ \bibnamefont {Mihaila}},
  \bibinfo {author} {\bibfnamefont {P.}~\bibnamefont {Marquard}}, \bibinfo
  {author} {\bibfnamefont {I.~F.}\ \bibnamefont {Herbut}},\ and\ \bibinfo
  {author} {\bibfnamefont {M.~M.}\ \bibnamefont {Scherer}},\ }\href
  {https://doi.org/10.1103/PhysRevD.96.096010} {\bibfield  {journal} {\bibinfo
  {journal} {Phys. Rev. D}\ }\textbf {\bibinfo {volume} {96}},\ \bibinfo
  {pages} {096010} (\bibinfo {year} {2017})}\BibitemShut {NoStop}%
\bibitem [{\citenamefont {Liu}\ \emph {et~al.}(2020)\citenamefont {Liu},
  \citenamefont {Wang}, \citenamefont {Sun},\ and\ \citenamefont
  {Meng}}]{Liu-Designer-2020}%
  \BibitemOpen
  \bibfield  {author} {\bibinfo {author} {\bibfnamefont {Y.}~\bibnamefont
  {Liu}}, \bibinfo {author} {\bibfnamefont {W.}~\bibnamefont {Wang}}, \bibinfo
  {author} {\bibfnamefont {K.}~\bibnamefont {Sun}},\ and\ \bibinfo {author}
  {\bibfnamefont {Z.~Y.}\ \bibnamefont {Meng}},\ }\href
  {https://doi.org/10.1103/PhysRevB.101.064308} {\bibfield  {journal} {\bibinfo
   {journal} {Phys. Rev. B}\ }\textbf {\bibinfo {volume} {101}},\ \bibinfo
  {pages} {064308} (\bibinfo {year} {2020})}\BibitemShut {NoStop}%
\bibitem [{\citenamefont {Liu}\ \emph {et~al.}(2019{\natexlab{a}})\citenamefont
  {Liu}, \citenamefont {Xu}, \citenamefont {Qi}, \citenamefont {Sun},\ and\
  \citenamefont {Meng}}]{Liu-Elective-2019}%
  \BibitemOpen
  \bibfield  {author} {\bibinfo {author} {\bibfnamefont {Z.~H.}\ \bibnamefont
  {Liu}}, \bibinfo {author} {\bibfnamefont {X.~Y.}\ \bibnamefont {Xu}},
  \bibinfo {author} {\bibfnamefont {Y.}~\bibnamefont {Qi}}, \bibinfo {author}
  {\bibfnamefont {K.}~\bibnamefont {Sun}},\ and\ \bibinfo {author}
  {\bibfnamefont {Z.~Y.}\ \bibnamefont {Meng}},\ }\href
  {https://doi.org/10.1103/PhysRevB.99.085114} {\bibfield  {journal} {\bibinfo
  {journal} {Phys. Rev. B}\ }\textbf {\bibinfo {volume} {99}},\ \bibinfo
  {pages} {085114} (\bibinfo {year} {2019}{\natexlab{a}})}\BibitemShut
  {NoStop}%
\bibitem [{\citenamefont {He}\ \emph {et~al.}(2018)\citenamefont {He},
  \citenamefont {Xu}, \citenamefont {Sun}, \citenamefont {Assaad},
  \citenamefont {Meng},\ and\ \citenamefont {Lu}}]{He-Dynamical-2018}%
  \BibitemOpen
  \bibfield  {author} {\bibinfo {author} {\bibfnamefont {Y.-Y.}\ \bibnamefont
  {He}}, \bibinfo {author} {\bibfnamefont {X.~Y.}\ \bibnamefont {Xu}}, \bibinfo
  {author} {\bibfnamefont {K.}~\bibnamefont {Sun}}, \bibinfo {author}
  {\bibfnamefont {F.~F.}\ \bibnamefont {Assaad}}, \bibinfo {author}
  {\bibfnamefont {Z.~Y.}\ \bibnamefont {Meng}},\ and\ \bibinfo {author}
  {\bibfnamefont {Z.-Y.}\ \bibnamefont {Lu}},\ }\href
  {https://doi.org/10.1103/PhysRevB.97.081110} {\bibfield  {journal} {\bibinfo
  {journal} {Phys. Rev. B}\ }\textbf {\bibinfo {volume} {97}},\ \bibinfo
  {pages} {081110} (\bibinfo {year} {2018})}\BibitemShut {NoStop}%
\bibitem [{\citenamefont {Litim}\ and\ \citenamefont
  {Zappal\`a}(2011)}]{litimIsing2011}%
  \BibitemOpen
  \bibfield  {author} {\bibinfo {author} {\bibfnamefont {D.~F.}\ \bibnamefont
  {Litim}}\ and\ \bibinfo {author} {\bibfnamefont {D.}~\bibnamefont
  {Zappal\`a}},\ }\href {https://doi.org/10.1103/PhysRevD.83.085009} {\bibfield
   {journal} {\bibinfo  {journal} {Phys. Rev. D}\ }\textbf {\bibinfo {volume}
  {83}},\ \bibinfo {pages} {085009} (\bibinfo {year} {2011})}\BibitemShut
  {NoStop}%
\bibitem [{\citenamefont {Kompaniets}\ and\ \citenamefont
  {Panzer}(2017)}]{kompanietsMinimally2017}%
  \BibitemOpen
  \bibfield  {author} {\bibinfo {author} {\bibfnamefont {M.~V.}\ \bibnamefont
  {Kompaniets}}\ and\ \bibinfo {author} {\bibfnamefont {E.}~\bibnamefont
  {Panzer}},\ }\href {https://doi.org/10.1103/PhysRevD.96.036016} {\bibfield
  {journal} {\bibinfo  {journal} {Phys. Rev. D}\ }\textbf {\bibinfo {volume}
  {96}},\ \bibinfo {pages} {036016} (\bibinfo {year} {2017})}\BibitemShut
  {NoStop}%
\bibitem [{\citenamefont {El-Showk}\ \emph {et~al.}(2012)\citenamefont
  {El-Showk}, \citenamefont {Paulos}, \citenamefont {Poland}, \citenamefont
  {Rychkov}, \citenamefont {Simmons-Duffin},\ and\ \citenamefont
  {Vichi}}]{showkSolving2012}%
  \BibitemOpen
  \bibfield  {author} {\bibinfo {author} {\bibfnamefont {S.}~\bibnamefont
  {El-Showk}}, \bibinfo {author} {\bibfnamefont {M.~F.}\ \bibnamefont
  {Paulos}}, \bibinfo {author} {\bibfnamefont {D.}~\bibnamefont {Poland}},
  \bibinfo {author} {\bibfnamefont {S.}~\bibnamefont {Rychkov}}, \bibinfo
  {author} {\bibfnamefont {D.}~\bibnamefont {Simmons-Duffin}},\ and\ \bibinfo
  {author} {\bibfnamefont {A.}~\bibnamefont {Vichi}},\ }\href
  {https://doi.org/10.1103/PhysRevD.86.025022} {\bibfield  {journal} {\bibinfo
  {journal} {Phys. Rev. D}\ }\textbf {\bibinfo {volume} {86}},\ \bibinfo
  {pages} {025022} (\bibinfo {year} {2012})}\BibitemShut {NoStop}%
\bibitem [{\citenamefont {Kos}\ \emph {et~al.}(2016)\citenamefont {Kos},
  \citenamefont {Poland}, \citenamefont {Simmons-Duffin},\ and\ \citenamefont
  {Vichi}}]{kosPrecision2016}%
  \BibitemOpen
  \bibfield  {author} {\bibinfo {author} {\bibfnamefont {F.}~\bibnamefont
  {Kos}}, \bibinfo {author} {\bibfnamefont {D.}~\bibnamefont {Poland}},
  \bibinfo {author} {\bibfnamefont {D.}~\bibnamefont {Simmons-Duffin}},\ and\
  \bibinfo {author} {\bibfnamefont {A.}~\bibnamefont {Vichi}},\ }\href
  {https://doi.org/10.1007/JHEP08(2016)036} {\bibfield  {journal} {\bibinfo
  {journal} {Journal of High Energy Physics}\ }\textbf {\bibinfo {volume}
  {2016}},\ \bibinfo {pages} {36} (\bibinfo {year} {2016})}\BibitemShut
  {NoStop}%
\bibitem [{\citenamefont {Chester}\ \emph {et~al.}(2020)\citenamefont
  {Chester}, \citenamefont {Landry}, \citenamefont {Liu}, \citenamefont
  {Poland}, \citenamefont {Simmons-Duffin}, \citenamefont {Su},\ and\
  \citenamefont {Vichi}}]{chesterCarving2020}%
  \BibitemOpen
  \bibfield  {author} {\bibinfo {author} {\bibfnamefont {S.~M.}\ \bibnamefont
  {Chester}}, \bibinfo {author} {\bibfnamefont {W.}~\bibnamefont {Landry}},
  \bibinfo {author} {\bibfnamefont {J.}~\bibnamefont {Liu}}, \bibinfo {author}
  {\bibfnamefont {D.}~\bibnamefont {Poland}}, \bibinfo {author} {\bibfnamefont
  {D.}~\bibnamefont {Simmons-Duffin}}, \bibinfo {author} {\bibfnamefont
  {N.}~\bibnamefont {Su}},\ and\ \bibinfo {author} {\bibfnamefont
  {A.}~\bibnamefont {Vichi}},\ }\href {https://doi.org/10.1007/JHEP06(2020)142}
  {\bibfield  {journal} {\bibinfo  {journal} {Journal of High Energy Physics}\
  }\textbf {\bibinfo {volume} {2020}},\ \bibinfo {pages} {142} (\bibinfo {year}
  {2020})}\BibitemShut {NoStop}%
\bibitem [{\citenamefont {Chester}\ \emph {et~al.}(2021)\citenamefont
  {Chester}, \citenamefont {Landry}, \citenamefont {Liu}, \citenamefont
  {Poland}, \citenamefont {Simmons-Duffin}, \citenamefont {Su},\ and\
  \citenamefont {Vichi}}]{chesterBootstrapping2021}%
  \BibitemOpen
  \bibfield  {author} {\bibinfo {author} {\bibfnamefont {S.~M.}\ \bibnamefont
  {Chester}}, \bibinfo {author} {\bibfnamefont {W.}~\bibnamefont {Landry}},
  \bibinfo {author} {\bibfnamefont {J.}~\bibnamefont {Liu}}, \bibinfo {author}
  {\bibfnamefont {D.}~\bibnamefont {Poland}}, \bibinfo {author} {\bibfnamefont
  {D.}~\bibnamefont {Simmons-Duffin}}, \bibinfo {author} {\bibfnamefont
  {N.}~\bibnamefont {Su}},\ and\ \bibinfo {author} {\bibfnamefont
  {A.}~\bibnamefont {Vichi}},\ }\href
  {https://doi.org/10.1103/PhysRevD.104.105013} {\bibfield  {journal} {\bibinfo
   {journal} {Phys. Rev. D}\ }\textbf {\bibinfo {volume} {104}},\ \bibinfo
  {pages} {105013} (\bibinfo {year} {2021})}\BibitemShut {NoStop}%
\bibitem [{\citenamefont {Hasenbusch}(2010)}]{hasenbuschFinite2010}%
  \BibitemOpen
  \bibfield  {author} {\bibinfo {author} {\bibfnamefont {M.}~\bibnamefont
  {Hasenbusch}},\ }\href {https://doi.org/10.1103/PhysRevB.82.174433}
  {\bibfield  {journal} {\bibinfo  {journal} {Phys. Rev. B}\ }\textbf {\bibinfo
  {volume} {82}},\ \bibinfo {pages} {174433} (\bibinfo {year}
  {2010})}\BibitemShut {NoStop}%
\bibitem [{\citenamefont {Gracey}(1990)}]{graceyThreeloop1990}%
  \BibitemOpen
  \bibfield  {author} {\bibinfo {author} {\bibfnamefont {J.}~\bibnamefont
  {Gracey}},\ }\href
  {https://doi.org/https://doi.org/10.1016/0550-3213(90)90186-H} {\bibfield
  {journal} {\bibinfo  {journal} {Nuclear Physics B}\ }\textbf {\bibinfo
  {volume} {341}},\ \bibinfo {pages} {403} (\bibinfo {year}
  {1990})}\BibitemShut {NoStop}%
\bibitem [{\citenamefont {Gracey}\ \emph {et~al.}(2016)\citenamefont {Gracey},
  \citenamefont {Luthe},\ and\ \citenamefont {Schr\"oder}}]{graceyFour2016}%
  \BibitemOpen
  \bibfield  {author} {\bibinfo {author} {\bibfnamefont {J.~A.}\ \bibnamefont
  {Gracey}}, \bibinfo {author} {\bibfnamefont {T.}~\bibnamefont {Luthe}},\ and\
  \bibinfo {author} {\bibfnamefont {Y.}~\bibnamefont {Schr\"oder}},\ }\href
  {https://doi.org/10.1103/PhysRevD.94.125028} {\bibfield  {journal} {\bibinfo
  {journal} {Phys. Rev. D}\ }\textbf {\bibinfo {volume} {94}},\ \bibinfo
  {pages} {125028} (\bibinfo {year} {2016})}\BibitemShut {NoStop}%
\bibitem [{\citenamefont {Vojta}\ \emph {et~al.}(2000)\citenamefont {Vojta},
  \citenamefont {Zhang},\ and\ \citenamefont {Sachdev}}]{vojtaQuantum2000}%
  \BibitemOpen
  \bibfield  {author} {\bibinfo {author} {\bibfnamefont {M.}~\bibnamefont
  {Vojta}}, \bibinfo {author} {\bibfnamefont {Y.}~\bibnamefont {Zhang}},\ and\
  \bibinfo {author} {\bibfnamefont {S.}~\bibnamefont {Sachdev}},\ }\href
  {https://doi.org/10.1103/PhysRevLett.85.4940} {\bibfield  {journal} {\bibinfo
   {journal} {Phys. Rev. Lett.}\ }\textbf {\bibinfo {volume} {85}},\ \bibinfo
  {pages} {4940} (\bibinfo {year} {2000})}\BibitemShut {NoStop}%
\bibitem [{\citenamefont {Schwab}\ \emph {et~al.}(2022)\citenamefont {Schwab},
  \citenamefont {Janssen}, \citenamefont {Sun}, \citenamefont {Meng},
  \citenamefont {Herbut}, \citenamefont {Vojta},\ and\ \citenamefont
  {Assaad}}]{schwabNematic2022}%
  \BibitemOpen
  \bibfield  {author} {\bibinfo {author} {\bibfnamefont {J.}~\bibnamefont
  {Schwab}}, \bibinfo {author} {\bibfnamefont {L.}~\bibnamefont {Janssen}},
  \bibinfo {author} {\bibfnamefont {K.}~\bibnamefont {Sun}}, \bibinfo {author}
  {\bibfnamefont {Z.~Y.}\ \bibnamefont {Meng}}, \bibinfo {author}
  {\bibfnamefont {I.~F.}\ \bibnamefont {Herbut}}, \bibinfo {author}
  {\bibfnamefont {M.}~\bibnamefont {Vojta}},\ and\ \bibinfo {author}
  {\bibfnamefont {F.~F.}\ \bibnamefont {Assaad}},\ }\href
  {https://doi.org/10.1103/PhysRevLett.128.157203} {\bibfield  {journal}
  {\bibinfo  {journal} {Phys. Rev. Lett.}\ }\textbf {\bibinfo {volume} {128}},\
  \bibinfo {pages} {157203} (\bibinfo {year} {2022})}\BibitemShut {NoStop}%
\bibitem [{\citenamefont {Geim}\ and\ \citenamefont
  {Novoselov}(2007)}]{geimRise2007}%
  \BibitemOpen
  \bibfield  {author} {\bibinfo {author} {\bibfnamefont {A.~K.}\ \bibnamefont
  {Geim}}\ and\ \bibinfo {author} {\bibfnamefont {K.~S.}\ \bibnamefont
  {Novoselov}},\ }\href {https://doi.org/10.1038/nmat1849} {\bibfield
  {journal} {\bibinfo  {journal} {Nature Materials}\ }\textbf {\bibinfo
  {volume} {6}},\ \bibinfo {pages} {183 } (\bibinfo {year} {2007})}\BibitemShut
  {NoStop}%
\bibitem [{\citenamefont {Herbut}(2006)}]{herbutInteractions2006}%
  \BibitemOpen
  \bibfield  {author} {\bibinfo {author} {\bibfnamefont {I.~F.}\ \bibnamefont
  {Herbut}},\ }\href {https://doi.org/10.1103/PhysRevLett.97.146401} {\bibfield
   {journal} {\bibinfo  {journal} {Phys. Rev. Lett.}\ }\textbf {\bibinfo
  {volume} {97}},\ \bibinfo {pages} {146401} (\bibinfo {year}
  {2006})}\BibitemShut {NoStop}%
\bibitem [{\citenamefont {Herbut}\ \emph
  {et~al.}(2009{\natexlab{a}})\citenamefont {Herbut}, \citenamefont
  {Juri\ifmmode \check{c}\else \v{c}\fi{}i\ifmmode~\acute{c}\else \'{c}\fi{}},\
  and\ \citenamefont {Roy}}]{herbutTheory2009}%
  \BibitemOpen
  \bibfield  {author} {\bibinfo {author} {\bibfnamefont {I.~F.}\ \bibnamefont
  {Herbut}}, \bibinfo {author} {\bibfnamefont {V.}~\bibnamefont {Juri\ifmmode
  \check{c}\else \v{c}\fi{}i\ifmmode~\acute{c}\else \'{c}\fi{}}},\ and\
  \bibinfo {author} {\bibfnamefont {B.}~\bibnamefont {Roy}},\ }\href
  {https://doi.org/10.1103/PhysRevB.79.085116} {\bibfield  {journal} {\bibinfo
  {journal} {Phys. Rev. B}\ }\textbf {\bibinfo {volume} {79}},\ \bibinfo
  {pages} {085116} (\bibinfo {year} {2009}{\natexlab{a}})}\BibitemShut
  {NoStop}%
\bibitem [{\citenamefont {Herbut}\ \emph
  {et~al.}(2009{\natexlab{b}})\citenamefont {Herbut}, \citenamefont
  {Juri\ifmmode \check{c}\else \v{c}\fi{}i\ifmmode~\acute{c}\else \'{c}\fi{}},\
  and\ \citenamefont {Vafek}}]{herbutRelativistic2009}%
  \BibitemOpen
  \bibfield  {author} {\bibinfo {author} {\bibfnamefont {I.~F.}\ \bibnamefont
  {Herbut}}, \bibinfo {author} {\bibfnamefont {V.}~\bibnamefont {Juri\ifmmode
  \check{c}\else \v{c}\fi{}i\ifmmode~\acute{c}\else \'{c}\fi{}}},\ and\
  \bibinfo {author} {\bibfnamefont {O.}~\bibnamefont {Vafek}},\ }\href
  {https://doi.org/10.1103/PhysRevB.80.075432} {\bibfield  {journal} {\bibinfo
  {journal} {Phys. Rev. B}\ }\textbf {\bibinfo {volume} {80}},\ \bibinfo
  {pages} {075432} (\bibinfo {year} {2009}{\natexlab{b}})}\BibitemShut
  {NoStop}%
\bibitem [{\citenamefont {Cao}\ \emph {et~al.}(2018{\natexlab{a}})\citenamefont
  {Cao}, \citenamefont {Fatemi}, \citenamefont {Fang}, \citenamefont
  {Watanabe}, \citenamefont {Taniguchi}, \citenamefont {Kaxiras},\ and\
  \citenamefont {Jarillo-Herrero}}]{caoUnconventional2018}%
  \BibitemOpen
  \bibfield  {author} {\bibinfo {author} {\bibfnamefont {Y.}~\bibnamefont
  {Cao}}, \bibinfo {author} {\bibfnamefont {V.}~\bibnamefont {Fatemi}},
  \bibinfo {author} {\bibfnamefont {S.}~\bibnamefont {Fang}}, \bibinfo {author}
  {\bibfnamefont {K.}~\bibnamefont {Watanabe}}, \bibinfo {author}
  {\bibfnamefont {T.}~\bibnamefont {Taniguchi}}, \bibinfo {author}
  {\bibfnamefont {E.}~\bibnamefont {Kaxiras}},\ and\ \bibinfo {author}
  {\bibfnamefont {P.}~\bibnamefont {Jarillo-Herrero}},\ }\href
  {https://doi.org/10.1038/nature26160} {\bibfield  {journal} {\bibinfo
  {journal} {Nature}\ }\textbf {\bibinfo {volume} {556}},\ \bibinfo {pages}
  {43} (\bibinfo {year} {2018}{\natexlab{a}})}\BibitemShut {NoStop}%
\bibitem [{\citenamefont {Cao}\ \emph {et~al.}(2018{\natexlab{b}})\citenamefont
  {Cao}, \citenamefont {Fatemi}, \citenamefont {Demir}, \citenamefont {Fang},
  \citenamefont {Tomarken}, \citenamefont {Luo}, \citenamefont
  {Sanchez-Yamagishi}, \citenamefont {Watanabe}, \citenamefont {Taniguchi},
  \citenamefont {Kaxiras} \emph {et~al.}}]{caoCorrelated2018}%
  \BibitemOpen
  \bibfield  {author} {\bibinfo {author} {\bibfnamefont {Y.}~\bibnamefont
  {Cao}}, \bibinfo {author} {\bibfnamefont {V.}~\bibnamefont {Fatemi}},
  \bibinfo {author} {\bibfnamefont {A.}~\bibnamefont {Demir}}, \bibinfo
  {author} {\bibfnamefont {S.}~\bibnamefont {Fang}}, \bibinfo {author}
  {\bibfnamefont {S.~L.}\ \bibnamefont {Tomarken}}, \bibinfo {author}
  {\bibfnamefont {J.~Y.}\ \bibnamefont {Luo}}, \bibinfo {author} {\bibfnamefont
  {J.~D.}\ \bibnamefont {Sanchez-Yamagishi}}, \bibinfo {author} {\bibfnamefont
  {K.}~\bibnamefont {Watanabe}}, \bibinfo {author} {\bibfnamefont
  {T.}~\bibnamefont {Taniguchi}}, \bibinfo {author} {\bibfnamefont
  {E.}~\bibnamefont {Kaxiras}}, \emph {et~al.},\ }\href
  {https://doi.org/10.1038/nature26154} {\bibfield  {journal} {\bibinfo
  {journal} {Nature}\ }\textbf {\bibinfo {volume} {556}},\ \bibinfo {pages}
  {80} (\bibinfo {year} {2018}{\natexlab{b}})}\BibitemShut {NoStop}%
\bibitem [{\citenamefont {Kennes}\ \emph {et~al.}(2021)\citenamefont {Kennes},
  \citenamefont {Claassen}, \citenamefont {Xian}, \citenamefont {Georges},
  \citenamefont {Millis}, \citenamefont {Hone}, \citenamefont {Dean},
  \citenamefont {Basov}, \citenamefont {Pasupathy},\ and\ \citenamefont
  {Rubio}}]{kennesMoire2021}%
  \BibitemOpen
  \bibfield  {author} {\bibinfo {author} {\bibfnamefont {D.~M.}\ \bibnamefont
  {Kennes}}, \bibinfo {author} {\bibfnamefont {M.}~\bibnamefont {Claassen}},
  \bibinfo {author} {\bibfnamefont {L.}~\bibnamefont {Xian}}, \bibinfo {author}
  {\bibfnamefont {A.}~\bibnamefont {Georges}}, \bibinfo {author} {\bibfnamefont
  {A.~J.}\ \bibnamefont {Millis}}, \bibinfo {author} {\bibfnamefont
  {J.}~\bibnamefont {Hone}}, \bibinfo {author} {\bibfnamefont {C.~R.}\
  \bibnamefont {Dean}}, \bibinfo {author} {\bibfnamefont {D.~N.}\ \bibnamefont
  {Basov}}, \bibinfo {author} {\bibfnamefont {A.~N.}\ \bibnamefont
  {Pasupathy}},\ and\ \bibinfo {author} {\bibfnamefont {A.}~\bibnamefont
  {Rubio}},\ }\href {https://doi.org/10.1038/s41567-020-01154-3} {\bibfield
  {journal} {\bibinfo  {journal} {Nature Physics}\ }\textbf {\bibinfo {volume}
  {17}},\ \bibinfo {pages} {155 } (\bibinfo {year} {2021})}\BibitemShut
  {NoStop}%
\bibitem [{\citenamefont {Andrei}\ \emph {et~al.}(2021)\citenamefont {Andrei},
  \citenamefont {Efetov}, \citenamefont {Jarillo-Herrero}, \citenamefont
  {MacDonald}, \citenamefont {Mak}, \citenamefont {Senthil}, \citenamefont
  {Tutuc}, \citenamefont {Yazdani},\ and\ \citenamefont
  {Young}}]{andreiMarvels2021}%
  \BibitemOpen
  \bibfield  {author} {\bibinfo {author} {\bibfnamefont {E.~Y.}\ \bibnamefont
  {Andrei}}, \bibinfo {author} {\bibfnamefont {D.~K.}\ \bibnamefont {Efetov}},
  \bibinfo {author} {\bibfnamefont {P.}~\bibnamefont {Jarillo-Herrero}},
  \bibinfo {author} {\bibfnamefont {A.~H.}\ \bibnamefont {MacDonald}}, \bibinfo
  {author} {\bibfnamefont {K.~F.}\ \bibnamefont {Mak}}, \bibinfo {author}
  {\bibfnamefont {T.}~\bibnamefont {Senthil}}, \bibinfo {author} {\bibfnamefont
  {E.}~\bibnamefont {Tutuc}}, \bibinfo {author} {\bibfnamefont
  {A.}~\bibnamefont {Yazdani}},\ and\ \bibinfo {author} {\bibfnamefont {A.~F.}\
  \bibnamefont {Young}},\ }\href {https://doi.org/10.1038/s41578-021-00284-1}
  {\bibfield  {journal} {\bibinfo  {journal} {Nature Reviews Materials}\
  }\textbf {\bibinfo {volume} {6}},\ \bibinfo {pages} {201 } (\bibinfo {year}
  {2021})}\BibitemShut {NoStop}%
\bibitem [{\citenamefont {Yin}\ \emph {et~al.}(2022)\citenamefont {Yin},
  \citenamefont {Lian},\ and\ \citenamefont {Hasan}}]{yinTopological2022}%
  \BibitemOpen
  \bibfield  {author} {\bibinfo {author} {\bibfnamefont {J.-X.}\ \bibnamefont
  {Yin}}, \bibinfo {author} {\bibfnamefont {B.}~\bibnamefont {Lian}},\ and\
  \bibinfo {author} {\bibfnamefont {M.~Z.}\ \bibnamefont {Hasan}},\ }\href
  {https://doi.org/10.1038/s41586-022-05516-0} {\bibfield  {journal} {\bibinfo
  {journal} {Nature}\ }\textbf {\bibinfo {volume} {612}},\ \bibinfo {pages}
  {647 } (\bibinfo {year} {2022})}\BibitemShut {NoStop}%
\bibitem [{\citenamefont {Kang}\ \emph {et~al.}(2020)\citenamefont {Kang},
  \citenamefont {Ye}, \citenamefont {Fang}, \citenamefont {You}, \citenamefont
  {Levitan}, \citenamefont {Han}, \citenamefont {Facio}, \citenamefont
  {Jozwiak}, \citenamefont {Bostwick}, \citenamefont {Rotenberg}, \citenamefont
  {Chan}, \citenamefont {McDonald}, \citenamefont {Graf}, \citenamefont
  {Kaznatcheev}, \citenamefont {Vescovo}, \citenamefont {Bell}, \citenamefont
  {Kaxiras}, \citenamefont {van~den Brink}, \citenamefont {Richter},
  \citenamefont {Prasad~Ghimire}, \citenamefont {Checkelsky},\ and\
  \citenamefont {Comin}}]{kangDirac2020}%
  \BibitemOpen
  \bibfield  {author} {\bibinfo {author} {\bibfnamefont {M.}~\bibnamefont
  {Kang}}, \bibinfo {author} {\bibfnamefont {L.}~\bibnamefont {Ye}}, \bibinfo
  {author} {\bibfnamefont {S.}~\bibnamefont {Fang}}, \bibinfo {author}
  {\bibfnamefont {J.-S.}\ \bibnamefont {You}}, \bibinfo {author} {\bibfnamefont
  {A.}~\bibnamefont {Levitan}}, \bibinfo {author} {\bibfnamefont
  {M.}~\bibnamefont {Han}}, \bibinfo {author} {\bibfnamefont {J.~I.}\
  \bibnamefont {Facio}}, \bibinfo {author} {\bibfnamefont {C.}~\bibnamefont
  {Jozwiak}}, \bibinfo {author} {\bibfnamefont {A.}~\bibnamefont {Bostwick}},
  \bibinfo {author} {\bibfnamefont {E.}~\bibnamefont {Rotenberg}}, \bibinfo
  {author} {\bibfnamefont {M.~K.}\ \bibnamefont {Chan}}, \bibinfo {author}
  {\bibfnamefont {R.~D.}\ \bibnamefont {McDonald}}, \bibinfo {author}
  {\bibfnamefont {D.}~\bibnamefont {Graf}}, \bibinfo {author} {\bibfnamefont
  {K.}~\bibnamefont {Kaznatcheev}}, \bibinfo {author} {\bibfnamefont
  {E.}~\bibnamefont {Vescovo}}, \bibinfo {author} {\bibfnamefont {D.~C.}\
  \bibnamefont {Bell}}, \bibinfo {author} {\bibfnamefont {E.}~\bibnamefont
  {Kaxiras}}, \bibinfo {author} {\bibfnamefont {J.}~\bibnamefont {van~den
  Brink}}, \bibinfo {author} {\bibfnamefont {M.}~\bibnamefont {Richter}},
  \bibinfo {author} {\bibfnamefont {M.}~\bibnamefont {Prasad~Ghimire}},
  \bibinfo {author} {\bibfnamefont {J.~G.}\ \bibnamefont {Checkelsky}},\ and\
  \bibinfo {author} {\bibfnamefont {R.}~\bibnamefont {Comin}},\ }\href
  {https://doi.org/10.1038/s41563-019-0531-0} {\bibfield  {journal} {\bibinfo
  {journal} {Nature Materials}\ }\textbf {\bibinfo {volume} {19}},\ \bibinfo
  {pages} {163 } (\bibinfo {year} {2020})}\BibitemShut {NoStop}%
\bibitem [{\citenamefont {Chandrasekharan}\ and\ \citenamefont
  {Li}(2013)}]{Chandrasekharan-Quanum-2013}%
  \BibitemOpen
  \bibfield  {author} {\bibinfo {author} {\bibfnamefont {S.}~\bibnamefont
  {Chandrasekharan}}\ and\ \bibinfo {author} {\bibfnamefont {A.}~\bibnamefont
  {Li}},\ }\href {https://doi.org/10.1103/PhysRevD.88.021701} {\bibfield
  {journal} {\bibinfo  {journal} {Phys. Rev. D}\ }\textbf {\bibinfo {volume}
  {88}},\ \bibinfo {pages} {021701} (\bibinfo {year} {2013})}\BibitemShut
  {NoStop}%
\bibitem [{\citenamefont {Ihrig}\ \emph {et~al.}(2018)\citenamefont {Ihrig},
  \citenamefont {Mihaila},\ and\ \citenamefont
  {Scherer}}]{Ihrig-Critical-2018}%
  \BibitemOpen
  \bibfield  {author} {\bibinfo {author} {\bibfnamefont {B.}~\bibnamefont
  {Ihrig}}, \bibinfo {author} {\bibfnamefont {L.~N.}\ \bibnamefont {Mihaila}},\
  and\ \bibinfo {author} {\bibfnamefont {M.~M.}\ \bibnamefont {Scherer}},\
  }\href {https://doi.org/10.1103/PhysRevB.98.125109} {\bibfield  {journal}
  {\bibinfo  {journal} {Phys. Rev. B}\ }\textbf {\bibinfo {volume} {98}},\
  \bibinfo {pages} {125109} (\bibinfo {year} {2018})}\BibitemShut {NoStop}%
\bibitem [{\citenamefont {Iliesiu}\ \emph {et~al.}(2018)\citenamefont
  {Iliesiu}, \citenamefont {Kos}, \citenamefont {Poland}, \citenamefont
  {Pufu},\ and\ \citenamefont {Simmons-Duffin}}]{Iliesiu-Bootstrapping-2018}%
  \BibitemOpen
  \bibfield  {author} {\bibinfo {author} {\bibfnamefont {L.}~\bibnamefont
  {Iliesiu}}, \bibinfo {author} {\bibfnamefont {F.}~\bibnamefont {Kos}},
  \bibinfo {author} {\bibfnamefont {D.}~\bibnamefont {Poland}}, \bibinfo
  {author} {\bibfnamefont {S.~S.}\ \bibnamefont {Pufu}},\ and\ \bibinfo
  {author} {\bibfnamefont {D.}~\bibnamefont {Simmons-Duffin}},\ }\href
  {https://doi.org/10.1007/JHEP01(2018)036} {\bibfield  {journal} {\bibinfo
  {journal} {Journal of High Energy Physics}\ }\textbf {\bibinfo {volume}
  {2018}},\ \bibinfo {pages} {36} (\bibinfo {year} {2018})}\BibitemShut
  {NoStop}%
\bibitem [{\citenamefont {Tabatabaei}\ \emph {et~al.}(2022)\citenamefont
  {Tabatabaei}, \citenamefont {Negari}, \citenamefont {Maciejko},\ and\
  \citenamefont {Vaezi}}]{Tabatabaei-Chiral-2022}%
  \BibitemOpen
  \bibfield  {author} {\bibinfo {author} {\bibfnamefont {S.~M.}\ \bibnamefont
  {Tabatabaei}}, \bibinfo {author} {\bibfnamefont {A.-R.}\ \bibnamefont
  {Negari}}, \bibinfo {author} {\bibfnamefont {J.}~\bibnamefont {Maciejko}},\
  and\ \bibinfo {author} {\bibfnamefont {A.}~\bibnamefont {Vaezi}},\ }\href
  {https://doi.org/10.1103/PhysRevLett.128.225701} {\bibfield  {journal}
  {\bibinfo  {journal} {Phys. Rev. Lett.}\ }\textbf {\bibinfo {volume} {128}},\
  \bibinfo {pages} {225701} (\bibinfo {year} {2022})}\BibitemShut {NoStop}%
\bibitem [{\citenamefont {Bonati}\ \emph {et~al.}(2023)\citenamefont {Bonati},
  \citenamefont {Franchi}, \citenamefont {Pelissetto},\ and\ \citenamefont
  {Vicari}}]{Bonati-Chiral-2023}%
  \BibitemOpen
  \bibfield  {author} {\bibinfo {author} {\bibfnamefont {C.}~\bibnamefont
  {Bonati}}, \bibinfo {author} {\bibfnamefont {A.}~\bibnamefont {Franchi}},
  \bibinfo {author} {\bibfnamefont {A.}~\bibnamefont {Pelissetto}},\ and\
  \bibinfo {author} {\bibfnamefont {E.}~\bibnamefont {Vicari}},\ }\href
  {https://doi.org/10.1103/PhysRevD.107.034507} {\bibfield  {journal} {\bibinfo
   {journal} {Phys. Rev. D}\ }\textbf {\bibinfo {volume} {107}},\ \bibinfo
  {pages} {034507} (\bibinfo {year} {2023})}\BibitemShut {NoStop}%
\bibitem [{\citenamefont {Xu}\ \emph {et~al.}(2019)\citenamefont {Xu},
  \citenamefont {Liu}, \citenamefont {Pan}, \citenamefont {Qi}, \citenamefont
  {Sun},\ and\ \citenamefont {Meng}}]{xuRevealing2019}%
  \BibitemOpen
  \bibfield  {author} {\bibinfo {author} {\bibfnamefont {X.~Y.}\ \bibnamefont
  {Xu}}, \bibinfo {author} {\bibfnamefont {Z.~H.}\ \bibnamefont {Liu}},
  \bibinfo {author} {\bibfnamefont {G.}~\bibnamefont {Pan}}, \bibinfo {author}
  {\bibfnamefont {Y.}~\bibnamefont {Qi}}, \bibinfo {author} {\bibfnamefont
  {K.}~\bibnamefont {Sun}},\ and\ \bibinfo {author} {\bibfnamefont {Z.~Y.}\
  \bibnamefont {Meng}},\ }\href {https://doi.org/10.1088/1361-648X/ab3295}
  {\bibfield  {journal} {\bibinfo  {journal} {Journal of Physics: Condensed
  Matter}\ }\textbf {\bibinfo {volume} {31}},\ \bibinfo {pages} {463001}
  (\bibinfo {year} {2019})}\BibitemShut {NoStop}%
\bibitem [{\citenamefont {Liu}\ \emph {et~al.}(2019{\natexlab{b}})\citenamefont
  {Liu}, \citenamefont {Pan}, \citenamefont {Xu}, \citenamefont {Sun},\ and\
  \citenamefont {Meng}}]{liuItinerant2019}%
  \BibitemOpen
  \bibfield  {author} {\bibinfo {author} {\bibfnamefont {Z.~H.}\ \bibnamefont
  {Liu}}, \bibinfo {author} {\bibfnamefont {G.}~\bibnamefont {Pan}}, \bibinfo
  {author} {\bibfnamefont {X.~Y.}\ \bibnamefont {Xu}}, \bibinfo {author}
  {\bibfnamefont {K.}~\bibnamefont {Sun}},\ and\ \bibinfo {author}
  {\bibfnamefont {Z.~Y.}\ \bibnamefont {Meng}},\ }\href
  {https://doi.org/10.1073/pnas.1901751116} {\bibfield  {journal} {\bibinfo
  {journal} {Proceedings of the National Academy of Sciences}\ }\textbf
  {\bibinfo {volume} {116}},\ \bibinfo {pages} {16760} (\bibinfo {year}
  {2019}{\natexlab{b}})}\BibitemShut {NoStop}%
\bibitem [{\citenamefont {Lang}\ and\ \citenamefont
  {L\"auchli}(2019)}]{langQuantum2019}%
  \BibitemOpen
  \bibfield  {author} {\bibinfo {author} {\bibfnamefont {T.~C.}\ \bibnamefont
  {Lang}}\ and\ \bibinfo {author} {\bibfnamefont {A.~M.}\ \bibnamefont
  {L\"auchli}},\ }\href {https://doi.org/10.1103/PhysRevLett.123.137602}
  {\bibfield  {journal} {\bibinfo  {journal} {Phys. Rev. Lett.}\ }\textbf
  {\bibinfo {volume} {123}},\ \bibinfo {pages} {137602} (\bibinfo {year}
  {2019})}\BibitemShut {NoStop}%
\bibitem [{\citenamefont {Nielsen}\ and\ \citenamefont
  {Ninomiya}(1981{\natexlab{a}})}]{nielsenAbsenceI1981}%
  \BibitemOpen
  \bibfield  {author} {\bibinfo {author} {\bibfnamefont {H.}~\bibnamefont
  {Nielsen}}\ and\ \bibinfo {author} {\bibfnamefont {M.}~\bibnamefont
  {Ninomiya}},\ }\href
  {https://doi.org/https://doi.org/10.1016/0550-3213(81)90361-8} {\bibfield
  {journal} {\bibinfo  {journal} {Nuclear Physics B}\ }\textbf {\bibinfo
  {volume} {185}},\ \bibinfo {pages} {20} (\bibinfo {year}
  {1981}{\natexlab{a}})}\BibitemShut {NoStop}%
\bibitem [{\citenamefont {Nielsen}\ and\ \citenamefont
  {Ninomiya}(1981{\natexlab{b}})}]{nielsenAbsenceII1981}%
  \BibitemOpen
  \bibfield  {author} {\bibinfo {author} {\bibfnamefont {H.}~\bibnamefont
  {Nielsen}}\ and\ \bibinfo {author} {\bibfnamefont {M.}~\bibnamefont
  {Ninomiya}},\ }\href
  {https://doi.org/https://doi.org/10.1016/0550-3213(81)90524-1} {\bibfield
  {journal} {\bibinfo  {journal} {Nuclear Physics B}\ }\textbf {\bibinfo
  {volume} {193}},\ \bibinfo {pages} {173} (\bibinfo {year}
  {1981}{\natexlab{b}})}\BibitemShut {NoStop}%
\bibitem [{\citenamefont {Koziol}\ \emph {et~al.}(2021)\citenamefont {Koziol},
  \citenamefont {Langheld}, \citenamefont {Kapfer},\ and\ \citenamefont
  {Schmidt}}]{Koziol-Quantum-2021}%
  \BibitemOpen
  \bibfield  {author} {\bibinfo {author} {\bibfnamefont {J.~A.}\ \bibnamefont
  {Koziol}}, \bibinfo {author} {\bibfnamefont {A.}~\bibnamefont {Langheld}},
  \bibinfo {author} {\bibfnamefont {S.~C.}\ \bibnamefont {Kapfer}},\ and\
  \bibinfo {author} {\bibfnamefont {K.~P.}\ \bibnamefont {Schmidt}},\ }\href
  {https://doi.org/10.1103/PhysRevB.103.245135} {\bibfield  {journal} {\bibinfo
   {journal} {Phys. Rev. B}\ }\textbf {\bibinfo {volume} {103}},\ \bibinfo
  {pages} {245135} (\bibinfo {year} {2021})}\BibitemShut {NoStop}%
\bibitem [{\citenamefont {{Da Liao}}\ \emph {et~al.}(2022)\citenamefont {{Da
  Liao}}, \citenamefont {{Xu}}, \citenamefont {{Meng}},\ and\ \citenamefont
  {{Qi}}}]{liaoCaution2022}%
  \BibitemOpen
  \bibfield  {author} {\bibinfo {author} {\bibfnamefont {Y.}~\bibnamefont {{Da
  Liao}}}, \bibinfo {author} {\bibfnamefont {X.~Y.}\ \bibnamefont {{Xu}}},
  \bibinfo {author} {\bibfnamefont {Z.~Y.}\ \bibnamefont {{Meng}}},\ and\
  \bibinfo {author} {\bibfnamefont {Y.}~\bibnamefont {{Qi}}},\ }\href
  {https://doi.org/10.48550/arXiv.2210.04272} {\bibfield  {journal} {\bibinfo
  {journal} {arXiv e-prints}\ ,\ \bibinfo {eid} {arXiv:2210.04272}} (\bibinfo
  {year} {2022})},\ \Eprint {https://arxiv.org/abs/2210.04272}
  {arXiv:2210.04272 [cond-mat.str-el]} \BibitemShut {NoStop}%
\bibitem [{\citenamefont {{Wang}}\ \emph {et~al.}(2022)\citenamefont {{Wang}},
  \citenamefont {{Assaad}},\ and\ \citenamefont
  {{Ulybyshev}}}]{wangValidity2022}%
  \BibitemOpen
  \bibfield  {author} {\bibinfo {author} {\bibfnamefont {Z.}~\bibnamefont
  {{Wang}}}, \bibinfo {author} {\bibfnamefont {F.}~\bibnamefont {{Assaad}}},\
  and\ \bibinfo {author} {\bibfnamefont {M.}~\bibnamefont {{Ulybyshev}}},\
  }\href {https://doi.org/10.48550/arXiv.2211.02960} {\bibfield  {journal}
  {\bibinfo  {journal} {arXiv e-prints}\ ,\ \bibinfo {eid} {arXiv:2211.02960}}
  (\bibinfo {year} {2022})},\ \Eprint {https://arxiv.org/abs/2211.02960}
  {arXiv:2211.02960 [cond-mat.str-el]} \BibitemShut {NoStop}%
\bibitem [{sup()}]{suppl}%
  \BibitemOpen
  \href@noop {} {\bibinfo  {journal} {In this Supplementary Material, we
  discuss in detail the concept and implementation of the elective-momentum
  ultra-size (EMUS)-QMC method. We also provide details on how to perform the
  stochastic data collapse to unbiasedly extract the critical exponents}\
  }\BibitemShut {NoStop}%
\bibitem [{\citenamefont {Pujari}\ \emph {et~al.}(2016)\citenamefont {Pujari},
  \citenamefont {Lang}, \citenamefont {Murthy},\ and\ \citenamefont
  {Kaul}}]{pujariInteraction2016}%
  \BibitemOpen
\bibfield  {journal} {  }\bibfield  {author} {\bibinfo {author} {\bibfnamefont
  {S.}~\bibnamefont {Pujari}}, \bibinfo {author} {\bibfnamefont {T.~C.}\
  \bibnamefont {Lang}}, \bibinfo {author} {\bibfnamefont {G.}~\bibnamefont
  {Murthy}},\ and\ \bibinfo {author} {\bibfnamefont {R.~K.}\ \bibnamefont
  {Kaul}},\ }\href {https://doi.org/10.1103/PhysRevLett.117.086404} {\bibfield
  {journal} {\bibinfo  {journal} {Phys. Rev. Lett.}\ }\textbf {\bibinfo
  {volume} {117}},\ \bibinfo {pages} {086404} (\bibinfo {year}
  {2016})}\BibitemShut {NoStop}%
\bibitem [{\citenamefont {Sato}\ \emph {et~al.}(2017)\citenamefont {Sato},
  \citenamefont {Hohenadler},\ and\ \citenamefont {Assaad}}]{satoDirac2017}%
  \BibitemOpen
  \bibfield  {author} {\bibinfo {author} {\bibfnamefont {T.}~\bibnamefont
  {Sato}}, \bibinfo {author} {\bibfnamefont {M.}~\bibnamefont {Hohenadler}},\
  and\ \bibinfo {author} {\bibfnamefont {F.~F.}\ \bibnamefont {Assaad}},\
  }\href {https://doi.org/10.1103/PhysRevLett.119.197203} {\bibfield  {journal}
  {\bibinfo  {journal} {Phys. Rev. Lett.}\ }\textbf {\bibinfo {volume} {119}},\
  \bibinfo {pages} {197203} (\bibinfo {year} {2017})}\BibitemShut {NoStop}%
\bibitem [{\citenamefont {Ouyang}\ and\ \citenamefont
  {Xu}(2021)}]{ouyangProjection2021}%
  \BibitemOpen
  \bibfield  {author} {\bibinfo {author} {\bibfnamefont {Y.}~\bibnamefont
  {Ouyang}}\ and\ \bibinfo {author} {\bibfnamefont {X.~Y.}\ \bibnamefont
  {Xu}},\ }\href {https://doi.org/10.1103/PhysRevB.104.L241104} {\bibfield
  {journal} {\bibinfo  {journal} {Phys. Rev. B}\ }\textbf {\bibinfo {volume}
  {104}},\ \bibinfo {pages} {L241104} (\bibinfo {year} {2021})}\BibitemShut
  {NoStop}%
\bibitem [{\citenamefont {Da~Liao}\ \emph
  {et~al.}(2022{\natexlab{a}})\citenamefont {Da~Liao}, \citenamefont {Xu},
  \citenamefont {Meng},\ and\ \citenamefont {Qi}}]{liaoDiracI2022}%
  \BibitemOpen
  \bibfield  {author} {\bibinfo {author} {\bibfnamefont {Y.}~\bibnamefont
  {Da~Liao}}, \bibinfo {author} {\bibfnamefont {X.~Y.}\ \bibnamefont {Xu}},
  \bibinfo {author} {\bibfnamefont {Z.~Y.}\ \bibnamefont {Meng}},\ and\
  \bibinfo {author} {\bibfnamefont {Y.}~\bibnamefont {Qi}},\ }\href
  {https://doi.org/10.1103/PhysRevB.106.075111} {\bibfield  {journal} {\bibinfo
   {journal} {Phys. Rev. B}\ }\textbf {\bibinfo {volume} {106}},\ \bibinfo
  {pages} {075111} (\bibinfo {year} {2022}{\natexlab{a}})}\BibitemShut
  {NoStop}%
\bibitem [{\citenamefont {Da~Liao}\ \emph
  {et~al.}(2022{\natexlab{b}})\citenamefont {Da~Liao}, \citenamefont {Xu},
  \citenamefont {Meng},\ and\ \citenamefont {Qi}}]{liaoDiracII2022}%
  \BibitemOpen
  \bibfield  {author} {\bibinfo {author} {\bibfnamefont {Y.}~\bibnamefont
  {Da~Liao}}, \bibinfo {author} {\bibfnamefont {X.~Y.}\ \bibnamefont {Xu}},
  \bibinfo {author} {\bibfnamefont {Z.~Y.}\ \bibnamefont {Meng}},\ and\
  \bibinfo {author} {\bibfnamefont {Y.}~\bibnamefont {Qi}},\ }\href
  {https://doi.org/10.1103/PhysRevB.106.115149} {\bibfield  {journal} {\bibinfo
   {journal} {Phys. Rev. B}\ }\textbf {\bibinfo {volume} {106}},\ \bibinfo
  {pages} {115149} (\bibinfo {year} {2022}{\natexlab{b}})}\BibitemShut
  {NoStop}%
\bibitem [{\citenamefont {Da~Liao}\ \emph
  {et~al.}(2022{\natexlab{c}})\citenamefont {Da~Liao}, \citenamefont {Xu},
  \citenamefont {Meng},\ and\ \citenamefont {Qi}}]{liaoDiracIII2022}%
  \BibitemOpen
  \bibfield  {author} {\bibinfo {author} {\bibfnamefont {Y.}~\bibnamefont
  {Da~Liao}}, \bibinfo {author} {\bibfnamefont {X.~Y.}\ \bibnamefont {Xu}},
  \bibinfo {author} {\bibfnamefont {Z.~Y.}\ \bibnamefont {Meng}},\ and\
  \bibinfo {author} {\bibfnamefont {Y.}~\bibnamefont {Qi}},\ }\href
  {https://doi.org/10.1103/PhysRevB.106.155159} {\bibfield  {journal} {\bibinfo
   {journal} {Phys. Rev. B}\ }\textbf {\bibinfo {volume} {106}},\ \bibinfo
  {pages} {155159} (\bibinfo {year} {2022}{\natexlab{c}})}\BibitemShut
  {NoStop}%
\bibitem [{\citenamefont {Liu}\ \emph {et~al.}(2019{\natexlab{c}})\citenamefont
  {Liu}, \citenamefont {Wang}, \citenamefont {Sato}, \citenamefont
  {Hohenadler}, \citenamefont {Wang}, \citenamefont {Guo},\ and\ \citenamefont
  {Assaad}}]{liuSuperconductivity2019}%
  \BibitemOpen
  \bibfield  {author} {\bibinfo {author} {\bibfnamefont {Y.}~\bibnamefont
  {Liu}}, \bibinfo {author} {\bibfnamefont {Z.}~\bibnamefont {Wang}}, \bibinfo
  {author} {\bibfnamefont {T.}~\bibnamefont {Sato}}, \bibinfo {author}
  {\bibfnamefont {M.}~\bibnamefont {Hohenadler}}, \bibinfo {author}
  {\bibfnamefont {C.}~\bibnamefont {Wang}}, \bibinfo {author} {\bibfnamefont
  {W.}~\bibnamefont {Guo}},\ and\ \bibinfo {author} {\bibfnamefont {F.~F.}\
  \bibnamefont {Assaad}},\ }\href
  {https://www.nature.com/articles/s41467-019-10372-0} {\bibfield  {journal}
  {\bibinfo  {journal} {Nature communications}\ }\textbf {\bibinfo {volume}
  {10}},\ \bibinfo {pages} {1} (\bibinfo {year}
  {2019}{\natexlab{c}})}\BibitemShut {NoStop}%
\bibitem [{\citenamefont {Wang}\ \emph {et~al.}(2021)\citenamefont {Wang},
  \citenamefont {Liu}, \citenamefont {Sato}, \citenamefont {Hohenadler},
  \citenamefont {Wang}, \citenamefont {Guo},\ and\ \citenamefont
  {Assaad}}]{wangDoping2021}%
  \BibitemOpen
  \bibfield  {author} {\bibinfo {author} {\bibfnamefont {Z.}~\bibnamefont
  {Wang}}, \bibinfo {author} {\bibfnamefont {Y.}~\bibnamefont {Liu}}, \bibinfo
  {author} {\bibfnamefont {T.}~\bibnamefont {Sato}}, \bibinfo {author}
  {\bibfnamefont {M.}~\bibnamefont {Hohenadler}}, \bibinfo {author}
  {\bibfnamefont {C.}~\bibnamefont {Wang}}, \bibinfo {author} {\bibfnamefont
  {W.}~\bibnamefont {Guo}},\ and\ \bibinfo {author} {\bibfnamefont {F.~F.}\
  \bibnamefont {Assaad}},\ }\href
  {https://doi.org/10.1103/PhysRevLett.126.205701} {\bibfield  {journal}
  {\bibinfo  {journal} {Phys. Rev. Lett.}\ }\textbf {\bibinfo {volume} {126}},\
  \bibinfo {pages} {205701} (\bibinfo {year} {2021})}\BibitemShut {NoStop}%
\bibitem [{\citenamefont {Da~Liao}\ \emph {et~al.}(2019)\citenamefont
  {Da~Liao}, \citenamefont {Meng},\ and\ \citenamefont {Xu}}]{liaoValence2019}%
  \BibitemOpen
  \bibfield  {author} {\bibinfo {author} {\bibfnamefont {Y.}~\bibnamefont
  {Da~Liao}}, \bibinfo {author} {\bibfnamefont {Z.~Y.}\ \bibnamefont {Meng}},\
  and\ \bibinfo {author} {\bibfnamefont {X.~Y.}\ \bibnamefont {Xu}},\ }\href
  {https://doi.org/10.1103/PhysRevLett.123.157601} {\bibfield  {journal}
  {\bibinfo  {journal} {Phys. Rev. Lett.}\ }\textbf {\bibinfo {volume} {123}},\
  \bibinfo {pages} {157601} (\bibinfo {year} {2019})}\BibitemShut {NoStop}%
\bibitem [{\citenamefont {Da~Liao}\ \emph {et~al.}(2021)\citenamefont
  {Da~Liao}, \citenamefont {Kang}, \citenamefont {Brei\o{}}, \citenamefont
  {Xu}, \citenamefont {Wu}, \citenamefont {Andersen}, \citenamefont
  {Fernandes},\ and\ \citenamefont {Meng}}]{liaoCorrelation2021}%
  \BibitemOpen
  \bibfield  {author} {\bibinfo {author} {\bibfnamefont {Y.}~\bibnamefont
  {Da~Liao}}, \bibinfo {author} {\bibfnamefont {J.}~\bibnamefont {Kang}},
  \bibinfo {author} {\bibfnamefont {C.~N.}\ \bibnamefont {Brei\o{}}}, \bibinfo
  {author} {\bibfnamefont {X.~Y.}\ \bibnamefont {Xu}}, \bibinfo {author}
  {\bibfnamefont {H.-Q.}\ \bibnamefont {Wu}}, \bibinfo {author} {\bibfnamefont
  {B.~M.}\ \bibnamefont {Andersen}}, \bibinfo {author} {\bibfnamefont {R.~M.}\
  \bibnamefont {Fernandes}},\ and\ \bibinfo {author} {\bibfnamefont {Z.~Y.}\
  \bibnamefont {Meng}},\ }\href {https://doi.org/10.1103/PhysRevX.11.011014}
  {\bibfield  {journal} {\bibinfo  {journal} {Phys. Rev. X}\ }\textbf {\bibinfo
  {volume} {11}},\ \bibinfo {pages} {011014} (\bibinfo {year}
  {2021})}\BibitemShut {NoStop}%
\bibitem [{\citenamefont {Liao}\ \emph {et~al.}(2021)\citenamefont {Liao},
  \citenamefont {Xu}, \citenamefont {Meng},\ and\ \citenamefont
  {Kang}}]{liaoCorrelationReview2021}%
  \BibitemOpen
  \bibfield  {author} {\bibinfo {author} {\bibfnamefont {Y.-D.}\ \bibnamefont
  {Liao}}, \bibinfo {author} {\bibfnamefont {X.-Y.}\ \bibnamefont {Xu}},
  \bibinfo {author} {\bibfnamefont {Z.-Y.}\ \bibnamefont {Meng}},\ and\
  \bibinfo {author} {\bibfnamefont {J.}~\bibnamefont {Kang}},\ }\href
  {https://doi.org/10.1088/1674-1056/abcfa3} {\bibfield  {journal} {\bibinfo
  {journal} {Chinese Physics B}\ }\textbf {\bibinfo {volume} {30}},\ \bibinfo
  {pages} {017305} (\bibinfo {year} {2021})}\BibitemShut {NoStop}%
\bibitem [{\citenamefont {Pan}\ \emph {et~al.}(2022{\natexlab{a}})\citenamefont
  {Pan}, \citenamefont {Zhang}, \citenamefont {Li}, \citenamefont {Sun},\ and\
  \citenamefont {Meng}}]{panDynamical2022}%
  \BibitemOpen
  \bibfield  {author} {\bibinfo {author} {\bibfnamefont {G.}~\bibnamefont
  {Pan}}, \bibinfo {author} {\bibfnamefont {X.}~\bibnamefont {Zhang}}, \bibinfo
  {author} {\bibfnamefont {H.}~\bibnamefont {Li}}, \bibinfo {author}
  {\bibfnamefont {K.}~\bibnamefont {Sun}},\ and\ \bibinfo {author}
  {\bibfnamefont {Z.~Y.}\ \bibnamefont {Meng}},\ }\href
  {https://doi.org/10.1103/PhysRevB.105.L121110} {\bibfield  {journal}
  {\bibinfo  {journal} {Phys. Rev. B}\ }\textbf {\bibinfo {volume} {105}},\
  \bibinfo {pages} {L121110} (\bibinfo {year}
  {2022}{\natexlab{a}})}\BibitemShut {NoStop}%
\bibitem [{\citenamefont {Pan}\ \emph {et~al.}(2022{\natexlab{b}})\citenamefont
  {Pan}, \citenamefont {Jiang},\ and\ \citenamefont {Meng}}]{panSport2022}%
  \BibitemOpen
  \bibfield  {author} {\bibinfo {author} {\bibfnamefont {G.}~\bibnamefont
  {Pan}}, \bibinfo {author} {\bibfnamefont {W.}~\bibnamefont {Jiang}},\ and\
  \bibinfo {author} {\bibfnamefont {Z.~Y.}\ \bibnamefont {Meng}},\ }\href
  {https://doi.org/10.1088/1674-1056/aca083} {\bibfield  {journal} {\bibinfo
  {journal} {Chinese Physics B}\ }\textbf {\bibinfo {volume} {31}},\ \bibinfo
  {pages} {127101} (\bibinfo {year} {2022}{\natexlab{b}})}\BibitemShut
  {NoStop}%
\bibitem [{\citenamefont {Xu}\ and\ \citenamefont
  {Grover}(2021)}]{xuCompeting2021}%
  \BibitemOpen
  \bibfield  {author} {\bibinfo {author} {\bibfnamefont {X.~Y.}\ \bibnamefont
  {Xu}}\ and\ \bibinfo {author} {\bibfnamefont {T.}~\bibnamefont {Grover}},\
  }\href {https://doi.org/10.1103/PhysRevLett.126.217002} {\bibfield  {journal}
  {\bibinfo  {journal} {Phys. Rev. Lett.}\ }\textbf {\bibinfo {volume} {126}},\
  \bibinfo {pages} {217002} (\bibinfo {year} {2021})}\BibitemShut {NoStop}%
\bibitem [{\citenamefont {Zhang}\ \emph {et~al.}(2022)\citenamefont {Zhang},
  \citenamefont {Sun}, \citenamefont {Li}, \citenamefont {Pan},\ and\
  \citenamefont {Meng}}]{zhangSuperconductivity2022}%
  \BibitemOpen
  \bibfield  {author} {\bibinfo {author} {\bibfnamefont {X.}~\bibnamefont
  {Zhang}}, \bibinfo {author} {\bibfnamefont {K.}~\bibnamefont {Sun}}, \bibinfo
  {author} {\bibfnamefont {H.}~\bibnamefont {Li}}, \bibinfo {author}
  {\bibfnamefont {G.}~\bibnamefont {Pan}},\ and\ \bibinfo {author}
  {\bibfnamefont {Z.~Y.}\ \bibnamefont {Meng}},\ }\href
  {https://doi.org/10.1103/PhysRevB.106.184517} {\bibfield  {journal} {\bibinfo
   {journal} {Phys. Rev. B}\ }\textbf {\bibinfo {volume} {106}},\ \bibinfo
  {pages} {184517} (\bibinfo {year} {2022})}\BibitemShut {NoStop}%
\bibitem [{\citenamefont {Jiang}\ \emph {et~al.}(2022)\citenamefont {Jiang},
  \citenamefont {Pan}, \citenamefont {Liu},\ and\ \citenamefont
  {Meng}}]{jiangSolving2022}%
  \BibitemOpen
  \bibfield  {author} {\bibinfo {author} {\bibfnamefont {W.}~\bibnamefont
  {Jiang}}, \bibinfo {author} {\bibfnamefont {G.}~\bibnamefont {Pan}}, \bibinfo
  {author} {\bibfnamefont {Y.}~\bibnamefont {Liu}},\ and\ \bibinfo {author}
  {\bibfnamefont {Z.-Y.}\ \bibnamefont {Meng}},\ }\href
  {https://doi.org/10.1088/1674-1056/ac4f52} {\bibfield  {journal} {\bibinfo
  {journal} {Chinese Physics B}\ }\textbf {\bibinfo {volume} {31}},\ \bibinfo
  {eid} {040504} (\bibinfo {year} {2022})}\BibitemShut {NoStop}%
\bibitem [{\citenamefont {Hirsch}(1985)}]{Hirsch_two_1985}%
  \BibitemOpen
  \bibfield  {author} {\bibinfo {author} {\bibfnamefont {J.~E.}\ \bibnamefont
  {Hirsch}},\ }\href {https://doi.org/10.1103/PhysRevB.31.4403} {\bibfield
  {journal} {\bibinfo  {journal} {Phys. Rev. B}\ }\textbf {\bibinfo {volume}
  {31}},\ \bibinfo {pages} {4403} (\bibinfo {year} {1985})}\BibitemShut
  {NoStop}%
\bibitem [{\citenamefont {Assaad}\ and\ \citenamefont
  {Evertz}(2008)}]{Assaad_world-line_2008}%
  \BibitemOpen
  \bibfield  {author} {\bibinfo {author} {\bibfnamefont {F.}~\bibnamefont
  {Assaad}}\ and\ \bibinfo {author} {\bibfnamefont {H.}~\bibnamefont
  {Evertz}},\ }\bibinfo {title} {World-line and determinantal quantum monte
  carlo methods for spins, phonons and electrons},\ in\ \href
  {https://doi.org/10.1007/978-3-540-74686-7_10} {\emph {\bibinfo {booktitle}
  {Computational Many-Particle Physics}}},\ \bibinfo {editor} {edited by\
  \bibinfo {editor} {\bibfnamefont {H.}~\bibnamefont {Fehske}}, \bibinfo
  {editor} {\bibfnamefont {R.}~\bibnamefont {Schneider}},\ and\ \bibinfo
  {editor} {\bibfnamefont {A.}~\bibnamefont {Wei{\ss}e}}}\ (\bibinfo
  {publisher} {Springer Berlin Heidelberg},\ \bibinfo {address} {Berlin,
  Heidelberg},\ \bibinfo {year} {2008})\ pp.\ \bibinfo {pages}
  {277--356}\BibitemShut {NoStop}%
\bibitem [{\citenamefont {Wang}\ \emph {et~al.}(2006)\citenamefont {Wang},
  \citenamefont {Beach},\ and\ \citenamefont {Sandvik}}]{Wang-High-2006}%
  \BibitemOpen
  \bibfield  {author} {\bibinfo {author} {\bibfnamefont {L.}~\bibnamefont
  {Wang}}, \bibinfo {author} {\bibfnamefont {K.~S.~D.}\ \bibnamefont {Beach}},\
  and\ \bibinfo {author} {\bibfnamefont {A.~W.}\ \bibnamefont {Sandvik}},\
  }\href {https://doi.org/10.1103/PhysRevB.73.014431} {\bibfield  {journal}
  {\bibinfo  {journal} {Phys. Rev. B}\ }\textbf {\bibinfo {volume} {73}},\
  \bibinfo {pages} {014431} (\bibinfo {year} {2006})}\BibitemShut {NoStop}%
\bibitem [{\citenamefont {{Yan}}\ \emph {et~al.}(2022)\citenamefont {{Yan}},
  \citenamefont {{Ran}}, \citenamefont {{Wang}}, \citenamefont {{Samajdar}},
  \citenamefont {{Rong}}, \citenamefont {{Sachdev}}, \citenamefont {{Qi}},\
  and\ \citenamefont {{Meng}}}]{yanFully2022}%
  \BibitemOpen
  \bibfield  {author} {\bibinfo {author} {\bibfnamefont {Z.}~\bibnamefont
  {{Yan}}}, \bibinfo {author} {\bibfnamefont {X.}~\bibnamefont {{Ran}}},
  \bibinfo {author} {\bibfnamefont {Y.-C.}\ \bibnamefont {{Wang}}}, \bibinfo
  {author} {\bibfnamefont {R.}~\bibnamefont {{Samajdar}}}, \bibinfo {author}
  {\bibfnamefont {J.}~\bibnamefont {{Rong}}}, \bibinfo {author} {\bibfnamefont
  {S.}~\bibnamefont {{Sachdev}}}, \bibinfo {author} {\bibfnamefont
  {Y.}~\bibnamefont {{Qi}}},\ and\ \bibinfo {author} {\bibfnamefont {Z.~Y.}\
  \bibnamefont {{Meng}}},\ }\href {https://doi.org/10.48550/arXiv.2205.04472}
  {\bibfield  {journal} {\bibinfo  {journal} {arXiv e-prints}\ ,\ \bibinfo
  {eid} {arXiv:2205.04472}} (\bibinfo {year} {2022})},\ \Eprint
  {https://arxiv.org/abs/2205.04472} {arXiv:2205.04472 [cond-mat.str-el]}
  \BibitemShut {NoStop}%
\end{thebibliography}%
\bibliographystyle{apsrev4-2}

\clearpage
\onecolumngrid

\begin{center}
	\textbf{\large Supplementary Material for "Emus live on the chiral Gross-Neveu-Yukawa archipelago"}
\end{center}
%\appendix
\setcounter{equation}{0}
\setcounter{figure}{0}
\setcounter{table}{0}
\setcounter{page}{1}
\makeatletter
\renewcommand{\theequation}{S\arabic{equation}}
\renewcommand{\thefigure}{S\arabic{figure}}
\setcounter{secnumdepth}{3}

\section{ EMUS-QMC}  
The EMUS-QMC~\cite{Liu-Elective-2019} scheme is different from the usual fermion determinant QMC (DQMC) simulation, in that, instead of simulating the whole system for a given size, here we focus on small patches of BZ that are important in the IR limit, and ignore the effect of those momenta far from these "hot spots". In the present model, the hot spots are the two Dirac points. Alternatively, one can think of EMUS working on a different model, whose dispersion near the Dirac points is same as the original one, and are cut off everywhere else. These two models, although have their differences in terms of non-universal observables, such as magnitude of order parameter, location of phase transition point etc., should have the same critical behaviour as they have the same IR structure. Such conclusion has been successfully shown in our previous works of EMUS-QMC on 2D Fermi surface coupled to quantum critical Ising bosons on square and triangular lattices~\cite{liuItinerant2019,xuRevealing2019,Liu-Elective-2019}.

To implement EMUS for the present model in Eq.~\eqref{eq:eq1}, we first divide bosons and fermions into two sub-lattices labeled $\alpha=1,2$, and perform Fourier transform to both fermionic and bosonic operators as,
\begin{equation}
\begin{cases}
\psi_{\vec{k},\alpha,\sigma}&=\frac{1}{\sqrt{N}}\sum_{i}\psi_{i,\alpha,\sigma}e^{-i\vec{k}\cdot\vec{r}_i},\\
\psi_{\vec{k},\alpha,\sigma}^\dagger&=\frac{1}{\sqrt{N}}\sum_{i}\psi_{i,\alpha,\sigma}^\dagger e^{i\vec{k}\cdot\vec{r}_i},\\
\phi_{\vec{q},\alpha}&=\frac{1}{\sqrt{N}}\sum_{i}\phi_{i,\alpha}e^{-i\vec{q}\cdot\vec{r}_i}.
\end{cases}
\end{equation}

One should note that, after Fourier transform, $\phi_{\vec{q},\alpha}$ is complex in the momentum space. A constrained on them is that all $\phi_{\vec{q},\alpha}$, $\phi_{-\vec{q},\alpha}$ pairs are complex conjugate to each other, so as to keep all real space bosonic fields $\phi_{i,\alpha}$ real scalers.

The fermion part of model is block-diagonalized into $2\times2$ matrixes describing hopping between two sub-lattices at each of the $N=L\times L$ $\vec{k}$-points
\begin{equation}
\begin{split}
H_f=\sum_{\vec{k},\sigma}[e^{i\frac{\pi}{4}}\cos(\frac{k_x+k_y}{2})+e^{-i\frac{\pi}{4}}\cos(\frac{k_x-k_y}{2})]\psi^\dagger_{k,2,\sigma}\psi_{k,1,\sigma}+h.c..
%$$H_f=e^{-i\frac{k_x-k_y}{2}}e^{i\frac{\pi}{4}}\cos(\frac{k_x+k_y}{2})+e^{i\frac{k_x+k_y}{2}}e^{-i\frac{\pi}{4}}\cos(\frac{k_x-k_y}{2})+h.c.$$
\end{split}
\end{equation}
The coupling term now becomes a summation over two momenta
\begin{equation}
\begin{split}
\label{eq-hc}
H_c=\sum_{\vec{k},\vec{k}',\sigma}[\cos(\frac{k_x+k'_x}{2})\phi_{\vec{k}-\vec{k}',2}-\cos(\frac{k_y+k'_y}{2})\phi_{\vec{k}-\vec{k}',1}]\psi_{\vec{k},1,\sigma}^\dagger\psi_{\vec{k}',1,\sigma}\\
+[\cos(\frac{k_y+k'_y}{2})\phi_{\vec{k}-\vec{k}',2}-\cos(\frac{k_x+k'_x}{2})\phi_{\vec{k}-\vec{k}',1}]\psi_{\vec{k},2,\sigma}^\dagger\psi_{\vec{k}',2,\sigma}.
\end{split}
\end{equation}

Compared with working in real space as the conventional DQMC scheme, there are three advantages in working in momentum space. 

First, one can now ignore the momentum point out of those patches as they relate to high energy excitation. In DQMC simulation, one needs to work with the fermion Green's function ${G(\vec{r},\vec{r}')}$, which is an ${2N\times 2N}$ matrix (for two spin flavors), where $N=L\times L$ is the number of primitive cells. In our EMUS scheme, we only keep track on two square region near Dirac points with size $\frac{1}{6}$ of BZ, and discretize those, that is the red grids in Fig.~\ref{fig:1} (c). We call the length of these grids $L_f$, and the full system size is $L=6L_f$. Note that $L_f$ takes only even number so that $X \text{ and }X'$ are on the grid, and the number of sites considered in each patch is ${N_f=(L_f+1)^2}\approx N/36$. Note that there is no periodic boundary condition in these small patches and hence the left and right sides are not identified. In the smallest possible system size $L_f=2$, one is actually simulating a $12\times 12$ system by considering only 9 momenta in one patch (see Fig.~\ref{fig:1} (d)).

Second, this cut-off in momentum space of fermions also induces a cut-off in that of bosons. Unlike the case in Ref.~\cite{Liu-Elective-2019}, here the bosonic system is ferromagnetic, hence has order vector $\Gamma=(0,0)$. As a result, the hopping between the two patches in fermion $\vec{k}$ space, although corresponds to low energy fermion excitations themselves, has bosonic field relating them of momentum far away from $\Gamma$, thus, is suppressed by the bosonic Hamiltonian. As a result, one only needs to consider momentum components relating $\vec{k}$ points within the same patch, that is the square region around $\Gamma$ with length $2L_f$,containing ${(2L_f+1)^2}$ sites.

Lastly, one can now perform the EMUS-QMC simulation fully on momentum space and "flip" the bosonic fields directly on their momentum components $\phi_{\vec{q}}$. When updating in the fermion determinant of DQMC, another important matrix is $\Delta=e^{-\Delta\tau H_c(\{\phi'\})}e^{\Delta\tau H_c(\{\phi\})}-\textbf{I}$, where ${\{\phi\}}$ (${\{\phi'\}}$) is the configuration of bosonic fields before (after) an update. Consider the Trotter decomposition $\Delta=e^{-\Delta\tau(H_c(\{\phi'\})-H_c(\{\phi\}))}-\textbf{I}+O(\Delta\tau^2)$. From Eq.~\eqref{eq-hc}, after flipping any $\phi_{\vec{q}}$, the changed entries in $H_c$ is the ones whose momenta differ by $\vec{q}$. In other words, $\Delta$ is not needed to save as a full $N_f^2$ matrix, it is block-diagonalized by groups of momenta that are related by $\vec{q}$.

Fig.~\ref{fig:1} (d) shows an example of patch size $L_f=2$. Suppose now the boson with momentum $\vec{q}=\frac{2\pi}{L}(1,2)$ is updated to be ${\phi_{\vec{q}}\ra\phi_{\vec{q}}+\Delta\phi}$, where ${\Delta\phi}$ is a random complex number with real part and imaginary part drawn from a uniform distribution ${[-C_{\vec{q}},C_{\vec{q}}]}$. Simultaneously, one needs to update $\phi_{-\vec{q}}$ with a complex conjugate change ${\phi_{\vec{-\vec{q}}}\ra\phi_{\vec{-\vec{q}}}+\Delta\phi^*}$. Then, for the fermion part, $\phi_{\pm\vec{q}}$ will relate two pairs of fermions, therefore the matrix $\Delta$ consists of two $2\times 2$ diagonal blocks. This can then be used to calculate the acceptance probability of fermion part $r_f=|\det(\textbf{I}+\Delta(\textbf{I}-G))|^2$, according to the usual DQMC scheme. That of boson part can be calculated by doing inverse Fourier transform on bosonic field to get its real space configuration back and calculate the full energy, finding out the energy difference, and the probability is $r_b=e^{-\Delta\tau\Delta E_b}$. The full acceptance probability is ${r=r_f\times r_b}$.

In practice, there are a few more things we did to improve the efficiency. 

(i) The amplitude of change in bosonic field ${C_{\vec{q}}}$ can be optimised such that it will not propose change that has acceptance rate too small, while large enough to make the update significant. 

(ii) ${C_{\vec{q}}}$ is larger when the updating momentum $\vec{q}$ is closer to $\Gamma$, because $\phi_{\vec{q}}$ should have larger magnitude on average. Together with (i), we have ${C_{\vec{q}}=ae^{-b|\vec{q}|}}$, where $a$ and $b$ are positive number to be optimise before simulation. An appropriate update strength leads to a shorter auto-correlation time~\cite{jiangSolving2022}, enabling us to achieve higher order of precision with same number of MC steps. 

(iii) Calculating the bosonic interaction ${(\phi_i-\phi_j)^2}$ in real space is time-consuming, as one needs to find all 16 neighbors for  $2N$ bosonic fields. Instead, one can express this term with their Fourier components, and only need to consider all ${\phi_{\vec{q}}\phi_{-\vec{q}}}$ pair interactions with appropriate coefficients. In this way, the time complexity in this part is significantly reduced. 

(iv) Due to the fact that the two Dirac points $X$ and ${X'}$ are equivalent by the rotational symmetry, $H_f$ and $H_c$ are identical in two patches. As a result, one only need to compute $\Delta$ in one of the patches and the overall fermion acceptance probability $r_f$ is the square of that calculated in one patch, in addition to the modulus square taken due to the two spin orientations and particle-hole symmetry.

\section{Determinant Quantum Monte Carlo (DQMC)}
Besides the aforementioned difference, the rest of the simulation follows the usual DQMC scheme\cite{Hirsch_two_1985,Assaad_world-line_2008}.

The fermion weight of each specific bosonic configuration is
$$\prod_{\sigma=\ua\da}\det(\textbf{I}+B^\sigma(\beta,0))$$
where the matrix $B$ is defined as
$$B^\sigma(l_2\Delta\tau,l_1\Delta\tau)=\prod_{l=l_1+1}^{l_2}e^{-\Delta\tau H_c(\{\phi\}_l)}e^{-\Delta\tau H_f}$$
and $H_c(\{\phi\}_l)$ is the coupling term depends on the bosonic fields configuration on the $l$-th imaginary time layer $\{\phi\}_l$.

When an update is proposed on the $l$-th layer, one has
\begin{align*}
&\det(\textbf{I}+B(\beta,l\Delta\tau)e^{-\Delta\tau H_c(\{\phi'\}_{l})}e^{\Delta\tau H_f}B((l-1)\Delta\tau,0))\\
=&\det(\textbf{I}+B(\beta,l\Delta\tau)e^{-\Delta\tau H_c(\{\phi'\}_{l})}e^{\Delta\tau H_c(\{\phi\}_{l})}B(l\Delta\tau,0))\\
=&\det(\textbf{I}+B(\beta,l\Delta\tau)(\textbf{I}+\Delta)B(l\Delta\tau,0))
\end{align*}
where the matrix $\Delta$ is defined as
$$\Delta=e^{-\Delta\tau H_c(\{\phi'\}_{l})}e^{\Delta\tau H_c(\{\phi\}_{l})}-\textbf{I}\approx e^{-\Delta\tau(H_c(\{\phi'\}_l)-H_c(\{\phi\}_l))}-\textbf{I}$$
as mentioned before. And the acceptance rate is calculated as the ratio of weights
\begin{align*}
&\frac{\det(\textbf{I}+B(\beta,l\Delta\tau)(\textbf{I}+\Delta)B(l\Delta\tau,0)}{\det(\textbf{I}+B(\beta,l\Delta\tau)B(l\Delta\tau,0))}\\
=&\frac{\det(\textbf{I}+B(\beta,0))+B(\beta,l\Delta\tau)\Delta B(l\Delta\tau,0)}{\det(\textbf{I}+B(\beta,0))}\\
=&\det(\textbf{I}+B(\beta,l\Delta\tau)\Delta B(l\Delta\tau,0)(\textbf{I}+B(\beta,0))^{-1})\\
=&\det(\textbf{I}+\Delta B(l\Delta\tau,0)(\textbf{I}+B(\beta,0))^{-1}B(\beta,l\Delta\tau))\\
=&\det(\textbf{I}+\Delta(\textbf{I}-(\textbf{I}+B(l\Delta\tau,0)B(\beta,l\Delta\tau))^{-1}))\\
=&\det(\textbf{I}+\Delta(\textbf{I}-G(l\Delta\tau,l\Delta\tau)))
\end{align*}
where the Sherman-Morrison formula is applied in the fourth equality and $G$ is the equal-time Green's function $G_{i,j}(\tau,\tau)=\langle \psi_{i}(\tau)\psi^\dagger_{j}(\tau)\rangle$.

Combining the two spin orientations and considering the particle-hole symmetry, one finally has
$$r_f=|\det(\textbf{I}+\Delta(\textbf{I}-G(l\Delta\tau,l\Delta\tau)))|^2$$

\section{Stochastic Data collapse}
For data collapse of order parameter shown in Fig.~\ref{fig:2} (c) and (d), we use a stochastic approach similar to that in Ref.~\cite{Wang-High-2006,yanFully2022}. During the data collapse, a line is fitted with the collapsed data and the goodness of collapse is associated to the goodness of fit. Here R squared value is used, which is widely used in statistics in evaluating variation between data and prediction. It is closer to 1 if the the model fits well with the data. It is defined as
$$R^2=1-\frac{S_{res}}{S_{tot}}=1-F\text{, where}$$
$$S_{res}=\sum_{i=1}^nw_i(y_i-\hat{y}_i)^2\text{,  }S_{tot}=\sum_{i=1}^nw_i(y_i-\bar{y})^2$$
and $F$ is the ratio we keep track of and would like to minimised. $\hat{y}_i$ is the $y$-values of the scaled data point, while $y_i$ is that of the fitting curve with $x$-value same as the $i$-th data point. $\bar{y}$ is the average value of $y_i$. $w_i$ are weights that have larger values near the critical point, and decay exponentially away from that. This parameter ensures the process focusing more on the portion inside the critical regime than those far away.
Basically, $S_{res}$ measures the variation between the rescaled data and the fitted value, and $S_{tot}$ measures that of the fitting curve itself.

Whenever a set of critical exponents need to be tested, we scale the data obtained from different system sizes according to the exponents, fit one curve to all these points, and then calculate the ratio $F$. If it is small, one can conclude that this set of exponents produce collapsed data that can be explained well by one single curve, hence the set of exponents is better in terms of data collapse.

\end{document}